\documentclass[arabic, epsf,twocolumn,showpacs,floatfix,preprintnumbers,amsmath,amssymb]{revtex4}
\usepackage{graphicx}
\input{epsf}
\epsfclipon

\begin{document} 

\title{Collective motion of self-propelled particles interacting without cohesion}

\author{Hugues Chat\'e}
\affiliation{CEA -- Service de Physique de l'\'Etat Condens\'e,
Centre d'Etudes de Saclay, 91191 Gif-sur-Yvette, France}

\author{Francesco Ginelli}
\affiliation{CEA -- Service de Physique de l'\'Etat Condens\'e,
Centre d'Etudes de Saclay, 91191 Gif-sur-Yvette, France}

\author{Guillaume Gr\'egoire} 
\affiliation{Mati\`ere et Syst\`emes Complexes,
CNRS UMR 7057, Universit\'e Paris--Diderot, Paris, France}

\author{Franck Raynaud}
\affiliation{CEA -- Service de Physique de l'\'Etat Condens\'e,
Centre d'Etudes de Saclay, 91191 Gif-sur-Yvette, France}
\affiliation{Mati\`ere et Syst\`emes Complexes,
CNRS UMR 7057, Universit\'e Paris--Diderot, Paris, France}

\date{\today}

\begin{abstract}
We present a comprehensive study of Vicsek-style self-propelled
particle models in two and three space dimensions. The onset of
collective motion in such stochastic models with only local alignment
interactions is studied in detail and shown to be discontinuous
(first-order like). The properties of the ordered, collectively moving
phase are investigated. In a large domain of parameter space including
the transition region, well-defined high-density and high-order
propagating solitary structures are shown to dominate the dynamics.
Far enough from the transition region, on the other hand, these
objects are not present. A statistically-homogeneous ordered phase is
then observed, which is characterized by anomalously-strong density
fluctuations, superdiffusion, and strong intermittency.
\end{abstract}

\maketitle

\section{Introduction}

Collective motion phenomena in nature have attracted the interest of
scientists and other authors for quite a long time~\cite{Pliny}.  The
question of the advantage of living and moving in groups, for
instance, is a favorite one among evolutionary
biologists~\cite{Parrish1997}.  In a different perspective, physicists
are mostly concerned with the mechanisms at the origin of collective
motion, especially when it manifests itself as a true, non-trivial,
emerging phenomenon, i.e. in the absence of some obvious cause like
the existence of a leader followed by the group, a strong geometrical
constraint forcing the displacement, or some external field or
gradient felt by the whole population.  Moreover, the ubiquity of the
phenomenon at all scales, from intra-cellular molecular cooperative
motion to the displacement in group of large animals, raises, for
physicists at least, the question of the existence of some universal
features possibly shared among many different situations.

One way of approaching these problems is to construct and study
minimal models of collective motion: if universal properties of
collective motion do exist, then they should appear clearly within
such models and thus could be efficiently determined there, before
being tested for in more elaborate models and real-world experiments
or observations.  Such is the underlying motivation of recent studies
of collective motion by a string of physicists~\cite{Albano,
Couzin2003, Czirok1997, Gregoire2001, Huepe, Toner1995}. Among them,
the group of Tamas Vicsek has put forward what is probably the
simplest possible model exhibiting collective motion in a non-trivial
manner.

In the ``Vicsek model''~\cite{Vicsek1995}, point particles 
move off-lattice at constant speed $v_{\rm 0}$, adjusting their direction of
motion to that of the average velocity of their neighbors, up to some
noise term accounting for external or internal perturbations (see
below for a precise definition).  For a finite density of particles in
a finite box, perfect alignment is reached easily in the absence of
noise: in this fluctuation-less collective motion, the macroscopic
velocity equals the microscopic one.  On the other hand, for
strong noise particles are essentially non-interacting random walkers
and their macroscopic velocity is zero, up to statistical fluctuations.

Vicsek {\it et al.} showed that the onset of collective motion occurs
at a finite noise level. In other words, there exists, in the
asymptotic limit, a fluctuating phase where the macroscopic velocity
of the total population is, on average, finite.  Working mostly in two
space dimensions, they concluded, on the basis of
numerical~\cite{Vicsek1995, Czirok1997} simulations, that the onset of
this ordered motion is well described as a novel non-equilibrium
continuous phase transition leading to long range order, at odds with
equilibrium where the continuous XY symmetry cannot be spontaneously
broken in two space dimensions and below~\cite{Mermin}.  This brought
support to the idea of universal properties of collective motion since
the scaling exponents and functions associated to such phase
transitions are expected to bear some degree of universality, even out
of equilibrium.

The above results caused a well-deserved stir and prompted a large
number of studies at various levels~\cite{Albano, Huepe, Bertin,
Birnir, Bussem, Chate2006, Couzin2002, Couzin2003, Couzin2005,
Csahok1995, Czirok1999, Czirok1999_2, Duparcmeur, Comment_gg,
Gregoire2001, Gregoire2003, Hemmingson, Levine, Mikhailov, Mogilner,
Oloan, dOrsogna, Ken, Simha2002, Simha2002_2, Szabo, Toner1995,
Toner1998, Toner1998_2, Topaz2004, Topaz2006, Vicsek1999}.  In
particular, two of us showed that the onset of collective motion is in
fact discontinuous~\cite{Gregoire2004}, and that the original
conclusion of Vicsek {\it et al.}  was based on numerical results
obtained at too small sizes~\cite{Vicsek1995,Czirok1997}.  More
recently, the discontinuous character of the transition was challenged
in two publications, one by the Vicsek group~\cite{Nagy} and another
by Aldana {\it et al.}~\cite{Aldana2007}.

Here, after a definition of the models involved (Section~\ref{models}),
we come back, in Section~\ref{transition}, to this central issue
and present a rather comprehensive study of the onset of collective motion
in Vicsek-style models. In Section~\ref{ordered},
we describe the ordered, collective motion phase. 
Section~\ref{discussion} is 
devoted to a general discussion of our results together with some perspectives.
Most of the numerical results shown were obtained in two space dimensions, 
but we also present three-dimensional results. 
Wherever no explicit mention is made, the default space dimension is two.
Similarly, the default boundary conditions are periodic in a square or cubic 
domain.

\section{The models}
\label{models}

\subsection{Vicsek model: angular noise}

Let us first recall the dynamical rule defining the 
Vicsek model~\cite{Vicsek1995}. 
Point particles labeled by an integer index $i$ move off-lattice 
in a space of dimension $d$
with a velocity $\vec{v_i}$ of fixed modulus $v_{\rm 0}=|\vec{v_i}|$.
The {\it direction} of motion of particle $i$ depends
on the average velocity of all particles (including $i$)
in the spherical neighbourhood $\mathcal{S}_i$ of radius $r_{\rm 0}$ centered on $i$.
The discrete-time dynamics is synchronous: the direction of
motion and the position of all particles are updated at each
timestep $\Delta t$, in a driven, overdamped manner:
\begin{equation}
\vec{v_i}(t+\Delta t) = v_{\rm 0} \,\, (\mathcal{R}_\eta\!\circ\vartheta)\left[
\sum_{j\in\mathcal{S}_i}\vec{v_j}(t)\right]
\label{EVicsek}
\end{equation}
where $\vartheta$ is a normalization operator
($\vartheta(\vec{w})=\vec{w}/|\vec{w}|$) and $\mathcal{R}_\eta$ performs a 
random rotation uniformely distributed around the argument vector:  
in $d=2$, $\mathcal{R}_\eta \vec{v}$ is uniformely distributed 
around $\vec{v}$ inside an arc of amplitude $ 2 \pi\,\eta$;
in $d=3$, it lies in the solid angle subtended by a spherical cap
of amplitude $4 \pi\,\eta$ and centered around  $\vec{v}$.
The particles positions $\vec{r_i}$ are then simply updated 
by streaming along the chosen direction as in
\begin{equation}
\vec{r_i}(t+\Delta t) =\vec{r_i}(t) + 
\Delta t \,\, \vec{v_i}(t+\Delta t) \;.
\label{Evol}
\end{equation}
Note that the original updating scheme proposed by Vicsek {\it et
al.} in~\cite{Vicsek1995} defined the speed as a backward difference,
although we are using a forward difference. The simpler updating
above, now adopted in most studies of Vicsek-style models, is not
expected to yield different results in the asymptotic limit of
infinite size and time.

\subsection{A different noise term: vectorial noise}

The ``angular'' noise term in the model defined above can be thought
of as arising from the errors committed when particles try to follow
the locally-averaged direction of motion.  One could argue on the
other hand that most of the randomness stems from the evaluation of
each interaction between particle $i$ and one of its neighbors,
because, e.g., of perception errors or turbulent fluctuations in the
medium.  This suggests to replace Eq.(\ref{EVicsek}) by:
\begin{equation}
\vec{v_i}(t+\Delta t) = v_{\rm 0} \, \vartheta\!\left[
\sum_{j\in\mathcal{S}_i}\!\vec{v_j}(t)
+\eta \mathcal{N}_i \vec{\xi} \right] 
\label{EN2}
\end{equation}
where $\vec{\xi}$ is a random unit vector and $\mathcal{N}_i$ is the 
number of particles in $\mathcal{S}_i$. 
It is easy to realize that this ``vectorial'' noise acts differently on the
system. While the intensity of angular noise is independent from the 
degree of local alignment, the influence of the vectorial noise decreases with
increasing local order.

\subsection{Repulsive force}
\label{repulsed}

In the original formulation of the Vicsek model as well as in the two
variants defined above, the only interaction is {\it alignment}.  In a
separate work~\cite{Gregoire2003}, we introduced a two-body
repulsion/attraction force, to account for the possibility of
maintaining the cohesion of a flock in an infinite space (something
the Vicsek model does not allow).  Here, we only study models without
cohesion. Nevertheless, we have considered, in the following, the case
of a pairwise repulsion ``force'', to estimate in particular the
possible influence of the absence of volume exclusion effects in the
basic model, which leaves the local density actually unbounded.  We
thus introduce the short ranged, purely repulsive interaction exerted by particle
$j$ on particle $i$:
\begin{equation}
\vec{f}_{ij} = -\vec{e}_{ij} \times 
\left[1+\exp(|\vec{r}_j-\vec{r}_i|/r_{\rm c}-2)\right]^{-1} \;,
\label{Eforce2}
\end{equation}  
where $\vec{e}_{ij}$ is the unit vector pointing from particle $i$ to $j$
and $r_{\rm c}<r_{\rm 0}$ is the typical repulsion range. 
Equations (\ref{EVicsek}) and (\ref{EN2}) are then respectively 
generalized to 
\begin{equation}
\vec{v_i}(t+\Delta t) = v_{\rm 0} \,\, (\mathcal{R}_\eta\!\circ\vartheta) \left[\sum_{j\in\mathcal{S}_i}\vec{v_j}(t) + 
\beta \sum_{j\in\mathcal{S}_i} \vec{f}_{ij}\right]
\label{EVicsek2}
\end{equation}
and
\begin{equation}
\vec{v_i}(t+\Delta t) = v_{\rm 0} \, \vartheta\!\left[
\sum_{j\in\mathcal{S}_i}\!\vec{v_j}(t) + \beta \sum_{j\in\mathcal{S}_i} \vec{f}_{ij}
+\eta \mathcal{N}_i \vec{\xi} \right]\;, 
\label{EN3}
\end{equation}
where $\beta$ measures the relative strength of repulsion with respect to 
alignment and noise strength. 

\subsection{Control and order parameters}

The natural order parameter for our polar particles 
is simply the macroscopic {\it mean velocity}, conveniently normalized by the 
microscopic velocity $v_{\rm 0}$
\begin{equation}
\vec{\varphi}(t)=\frac{1}{v_{\rm 0}}\langle \vec{v_i}(t)\rangle_i \,,
\label{globalop}
\end{equation}
where $\langle\cdot\rangle_i$ stands for the average over the 
whole population. Here, we mostly consider its modulus
$\varphi(t)=\left|\vec{\varphi}(t)\right|$, the {\it scalar order parameter}.

In the following, we set, without loss of generality,
$\Delta t = 1$ and $r_{\rm 0}=1$, and express all time- and length-scales
in terms of these units. Moreover repulsive force will be
studied by fixing $r_{\rm c} = 0.127$ and $\beta = 2.5$.

This leaves us with two main parameters for these models: the noise
amplitude $\eta$ and the global density of particles $\rho$. Recently, the
microscopic velocity $v_{\rm 0}$ has been argued to play a major role as
well~\cite{Nagy}. All three parameters ($\eta$, $\rho$, and $v_{\rm 0}$) are
considered below.

\section{Order-disorder transition at the onset of collective motion}
\label{transition}

\begin{figure}
\includegraphics[width=8.6cm,clip]{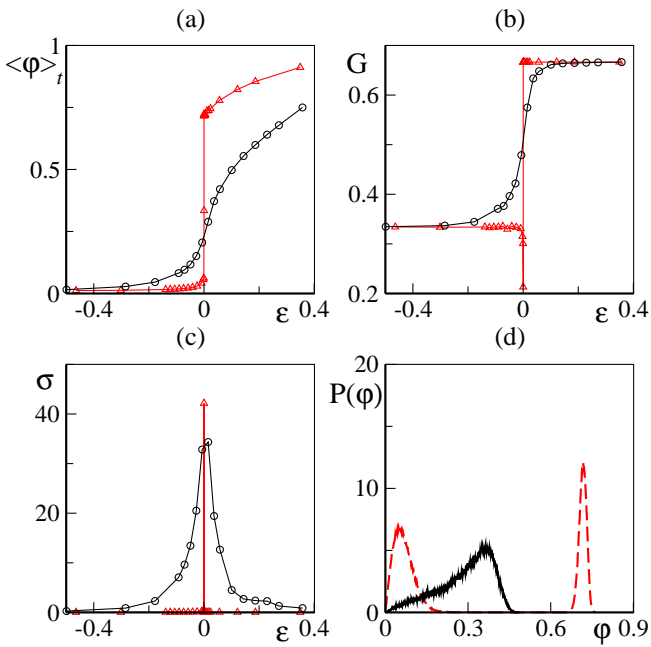}
\caption{
\label{fcompare}
(color online) Typical behavior across the onset of collective motion
for moderate size models ($\rho=2$, $v_{\rm 0}=0.5$, $L=64$) with angular
noise (black circles) and vectorial noise (red triangles). The reduced
noise amplitude $\varepsilon = 1 - \eta/\eta_{\rm t}$ is shown in abscissas
(transition points estimated at $\eta_{\rm t} = 0.6144(2)$ -- vectorial
noise -- and  $\eta_{\rm t} =0.478(5) $ -- angular noise).
(a) Time-averaged order parameter $\langle\varphi(t)\rangle_t$.
(b) Binder cumulant G.
(c) Variance $\sigma$ of $\varphi$;
(d) Order parameter distribution function $P$ at the transition point.
Bimodal distribution for vectorial noise dynamics (red dashed line),
unimodal shape for angular noise (black solid line). Time averages
have been computed over $3 \cdot 10^5$ timesteps.}
\end{figure}

As mentioned above, the original Vicsek model attracted a lot of
attention mostly because of the conclusions drawn from the early
numerical studies~\cite{Vicsek1995,Czirok1997}: the onset of
collective motion was found to be a novel continuous phase transition
spontaneously breaking rotational symmetry. However, it was later
shown in~\cite{Gregoire2004} that beyond the typical sizes considered
originally the {\it discontinuous} nature of the transition emerges,
irrespective of the form of the noise term. Recently, the
discontinuous character of the transition was argued to disappear in
the limit of small $v_{\rm 0}$~\cite{Nagy}.  We now address the problem of
the nature of the transition in full detail.

\begin{figure}[t]
\includegraphics[width=8.6cm,clip]{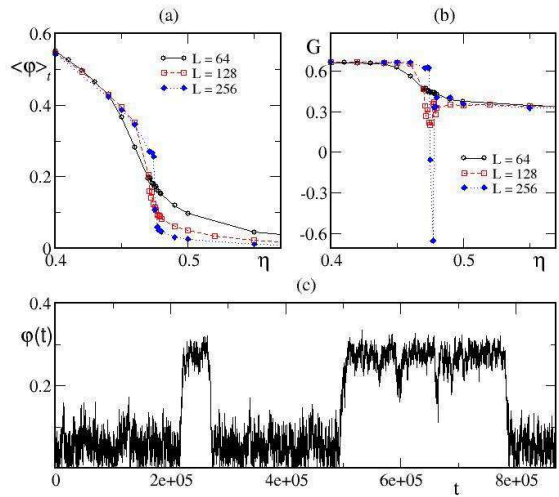}
\caption{\label{fig2}
(color online) FSS analysis of angular noise dynamics
($\rho=2$, $v_{\rm 0}=0.5$, time averages computed over $2 \cdot 10^7$ timesteps).
Time-averaged order parameter (a)
and Binder cumulant (b) as a function of noise for various system sizes $L$. 
(c) Piece of an order parameter time series close to the transition point 
($L=256$, $\eta=0.476$).}
\end{figure}

Even though there is no rigorous theory for finite-size-scaling (FSS)
for out-of-equilibrium phase transitions, there exists now ample
evidence that one can safely rely on the knowledge gained in
equilibrium systems~\cite{Lubeck, Marcq, Marcq2}. The FSS
approach~\cite{Binder,Privman} involves the numerical estimation of
various moments of the order parameter distribution as the linear
system size $L$ is systematically varied.  Of particular interest are
the variance
$$\sigma(\eta, L)=L^d\left(\langle\varphi^2\rangle_t -
\langle\varphi\rangle_t^2\right)$$
and the so-called Binder cumulant
\begin{equation}
G(\eta, L)= 1-\frac{\langle\varphi^4\rangle_t}{3\langle\varphi^2\rangle_t^2} \;,
\label{EBinder}
\end{equation}
where $\langle\cdot\rangle_t$ indicates time average.  The Binder
cumulant is especially useful in the case of continuous phase
transitions, because it is one of the simplest ratio of moments which
takes a universal value at the critical point $\eta_{\rm t}$, where all the
curves $G(\eta, L)$, obtained at different system sizes $L$, cross
each other.  At a first-order transition point, on the other hand, the
Binder cumulant exhibits a sharp drop towards negative
values~\cite{Binder1997}. This minimum is due to the simultaneous
contributions of the two phases coexisting at threshold.  Moreover, it
is easy to compute that $G(\eta, L) \approx 2/3$ in the ordered phase,
while for a disordered state with a continuous rotational simmetry one
has $G(\eta, L) \approx 1/3$ in $d=2$ and $G(\eta, L) \approx 4/9$ in
$d=3$. 
\begin{figure}
\includegraphics[width=8.6cm,clip]{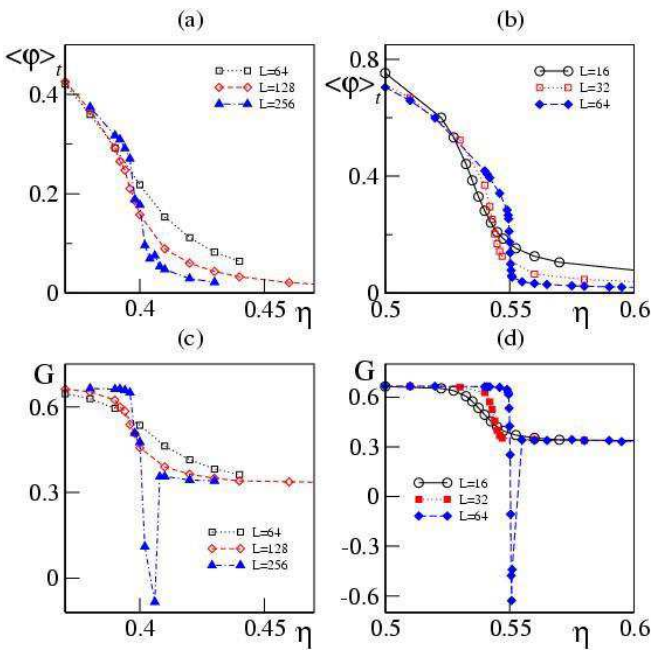}
\caption{\label{ffssvm2}
(color online)
Transition to collective motion with short range repulsive interactions.
Left panels: angular noise. Right panels: vectorial noise. 
(a,b): order parameter vs noise amplitude at different system sizes.
(c,d): Binder cumulant $G$ as a function of noise amplitude.
($\rho=2$, $v_{\rm 0}=0.3$, time averages carried over $10^7$ timesteps).}
\end{figure}

\subsection{Overture}
\label{overture}

As an overture, we analyze systems of moderate size 
in two dimensions ($N\approx 10^4$
particles) at the density $\rho=2$, typical of the initial studies
by Vicsek {\it et al.}, but with the slightly modified update rule 
(\ref{Evol}) and
for both angular and vectorial noise. The microscopic
velocity is set to $v_{\rm 0}=0.5$.

For angular noise, the transition looks indeed continuous, as found by
Vicsek {\it et al.}. On the other hand, the time averaged scalar order
parameter $\langle\varphi\rangle_t$ displays a sharp drop for
vectorial noise, and the Binder cumulant exibits a minimum at the
transition point, indicating a discontinuous phase transition
(Fig.~\ref{fcompare}a-b).  Simultaneously, the variance is almost
delta-peaked.  The difference betwen the two cases is also recorded in
the probability distribution function (PDF) of $\varphi$ which is
bimodal (phase coexistence) in the vectorial noise case
(Fig.~\ref{fcompare}c-d).

\begin{figure}
\includegraphics[width=8.6cm,clip]{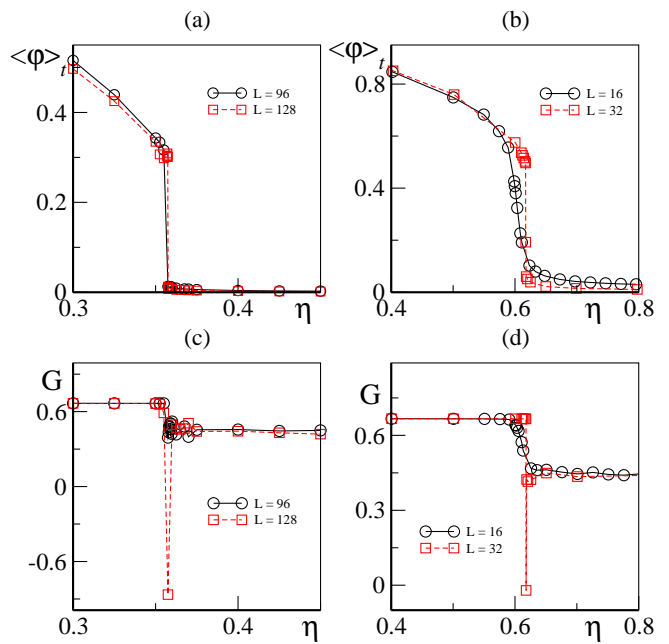}
\caption{\label{fig3D}
(color online)
Transition to collective motion in three spatial dimensions.
Left panels: angular noise. Right panels: vectorial noise. 
(a,b): time-averaged order parameter vs. noise amplitude at different
system sizes.
(c,d): Binder cumulant $G$ as a function of noise amplitude.
($\rho=0.5$, $v_{\rm 0}=0.5$, time averages carried over $10^5$ timesteps).}
\end{figure}

The qualitative difference observed upon changing the way noise is
implemented in the dynamics is, however, only a finite-size effect.
As shown in~\cite{Gregoire2004}, the transition in the angular noise
case reveals its asymptotic discontinuous character provided
large-enough system sizes $L$ are considered (Fig.~\ref{fig2}a-b).
Remaining for now at a qualitative level, we show in Fig.~\ref{fig2}c
a typical time-series of the order parameter for the angular noise
case in a large system in the transition region.  The sudden jumps
from the disordered phase to the ordered one and vice-versa are
evidence for metastability and phase coexistence.

Note that the system size beyond which the transition reveals its
discontinuous character for the angular noise case at density $\rho=2$
and velocity $v_{\rm 0}=0.5$ --- the conditions of the original papers
by Vicsek {\it et al.}--- is of the order of $L=128$, the maximum size
then considered. It is clear also from Fig.~\ref{fcompare} that the
discontinuous nature of the transition appears earlier, when
increasing system size, for vectorial noise than for angular noise.
Thus, finite-size effects are stronger for angular noise.  The same is
true when one is in presence of the repulsive interactions
(Fig.\ref{ffssvm2}).  Finally, the same scenario holds in three space
dimensions, with a {\it discontinous} phase transition separating the
ordered from the disordered phases for both angular and vectorial
noise (Fig.~\ref{fig3D}).

Before proceeding to a study of the complete phase diagram,
we detail now how a comprehensive FSS study can be performed on a particular case.

\subsection{Complete FSS analysis}
\label{FSSstudy}

\begin{figure}
\includegraphics[width=8.6cm,clip]{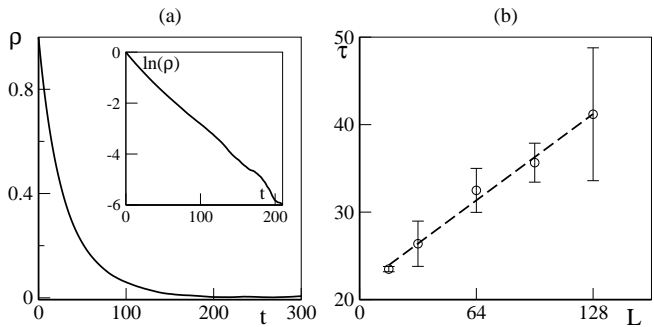}
\caption{\label{ftau} Correlation time $\tau$ of the order parameter
near the transition point for vectorial noise dynamics with
repulsion. System parameters are $\rho=2$, $v_{\rm 0}=0.5$ and $\eta \approx
\eta_{\rm t}$.  (a) Time correlation function $C(t)$ at $L=128$. The lin-log
inset shows the exponential decay.  (b) Correlation time $\tau$ as a
function of system size $L$.  The dashed line marks linear growth with
$L$.  Correlation functions were computed on samples of $\approx 10^6$
realizations for typically $10^3$ timesteps.}
\end{figure}

For historical reasons, the following study has been performed on the model 
with vectorial noise and repulsive force (Eq.~\ref{EN3}).
It has not been repeated in the simpler case of the ``pure'' 
Vicsek model because
its already high numerical cost would have been prohibitive due to the strong finite size effects.

As a first step, we estimated the correlation time $\tau(L)$, whose
knowledge is needed to control the quality of time-averaging: the
duration $T$ of numerical simulations has been taken much larger than
$\tau(L)$ ($T=100\tau$ in the largest systems, but typically
$10000\tau$ for smaller sizes). Moreover, $\tau$ is also useful to
correctly estimate the statistical errors on the various moments (as
$\langle\varphi\rangle_t$, $\sigma$, and $G$ ) of the PDF of the order
parameter, for which we used the Jackknife procedure~\cite{Efron}.
The correlation time was estimated near the transition (where it is
expected to be largest) as function of system size $L$ measuring the
exponential decay rate of the correlation function (Fig.~\ref{ftau}a)
\begin{equation}
\label{tcorr}
C(t)=
\langle\varphi(t_{\rm 0})\varphi(t_{\rm 0}+t)\rangle_t-\langle\varphi(t_{\rm 0})\rangle_{t_{\rm 0}}^2
\sim \exp{\left(-\frac{t}{\tau}\right)} \;.
\end{equation}
We found $\tau$ to vary roughly linearly with $L$ (see Fig.~\ref{ftau}b).
It is interesting to observe that at equilibrium, one would expect $\tau$
to scale as~\cite{Zinn}
$$
\tau=L^{\frac{d}{2}}\exp (\kappa L^{d-1} )
$$
where $\kappa$ is the surface tension of the metastable state.
Therefore, our result implies a very small or vanishing surface tension $\kappa\ll 1/L$,
a situation reminiscent of observations made in the 
cohesive case~\cite{Gregoire2003}, where the surface tension of a 
cohesive droplet was found to vanish near the onset.\\

Following  Borgs and Koteck\`y~\cite{Borgs}, the asymptotic 
coexistence point $\eta_{\rm t}$ (i.e. the first order transition point) 
can be determined from the asymptotic convergence
of various moments of the order parameter PDF. 
First, the observed discontinuity in $\langle\varphi(t)\rangle_t$,
located at $\eta_\varphi(L)$,
is expected to converge exponentially to $\eta_{\rm t}$ with $L$.
Second, the location of the susceptibility peak  $\eta_\sigma(L)$
--- which is the same as the 
peak in $\sigma$ provided some fluctuation-dissipation relation holds, 
see the Appendix --- also converges to  $\eta_{\rm t}$, 
albeit algebraically with an exponent $\gamma_{\sigma}$.
Third, the location of the minimum of $G$,  $\eta_G(L)$, is also expected to
converge algebraically to $\eta_{\rm t}$ with with an exponent
$\gamma_{G}=\gamma_{\sigma}$.

\begin{figure}
\includegraphics[width=8.6cm,clip]{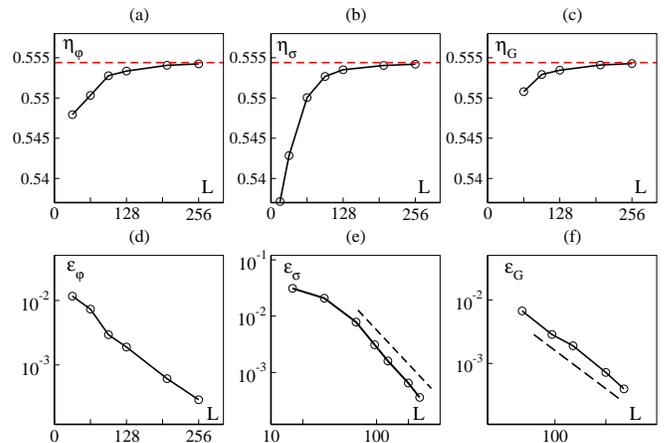}
\caption{(color online) FSS analysis of vectorial dynamics with short range repulsive
force ($\rho=2$, $v_{\rm 0}=0.3$).  Convergence of the finite-size
transition points measured from different moments of the order
parameter FSS to the asymptotic transition point $\eta_{\rm t}$ (see
fig.~\ref{ffssvm2}).  Upper panels: finite size transition points
estimated from (a) time average, (b) variance and (c) Binder
cumulant. The horizontal dashed line marks the estimated asymptotic
threshold $\eta_{\rm t}=0.5544(1)$.  Lower panels: scaling of the
finite size reduced noise $\varepsilon = 1 - \eta/\eta_{\rm t}$ transition
point. (d) Exponential convergence for the jump location in the time-averaged order parameter.  (e) Power-law
behavior of the variance peak position. (f) Power-law behavior of the Binder
cumulant minimum.  The dashed lines in (e-f) mark the estimated exponents
$\gamma_{\sigma}=\gamma_{G} = 2$.}
\label{fdisc}
\end{figure}

Interestingly, the value taken by these exponents actually depends on
the number of phases and of the dimension $d$ of the system: for
two-phase coexistence one has $\gamma_{G}=\gamma_{\sigma}=2d$, while
for more than two phases $\gamma_{G}=\gamma_{\sigma}=d$.  In
Figure~\ref{fdisc}, we show that our data are in good agreement with
all these predictions.  The three estimates of $\eta_{\rm t}$ are
consistent with each other within numerical accuracy. Moreover
$\eta_\varphi(L)$ is found to converge exponentially to the transitional
noise amplitude, while both $\eta_\sigma(L)$ and $\eta_G(L)$ show
algebraic convergence with an exponent close to 2. This agrees with
the fact that due to the continuous rotational symmetry, the ordered
phase is degenerate and amounts to an infinite number of possible
phases.

\subsection{Hysteresis}

\begin{figure}
\includegraphics[width=8.6cm,clip]{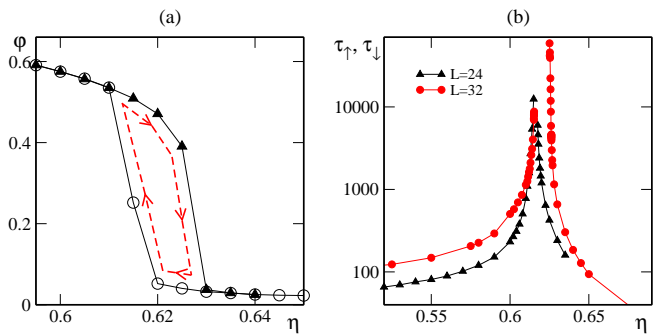}
\caption{(color online)
Hysteresis in three spatial dimensions with vectorial noise.
(a) order parameter vs noise strength along the hysteresis loop observed 
with a ramp rate of $2 \cdot 10^{-6}$ per time step ($\rho=1/2$, $v_{\rm 0}=0.5$, $L=32$).
Empty circles mark the path along the adiabatic increase of noise amplitude, full triangles for
adiabatic decrease.
(b) nucleation times from the disordered phase to the ordered phase 
($\tau_{\uparrow}$, left curves) and 
vice-versa ($\tau_{\downarrow}$, right curves) for two system sizes
(other parameters as in (a)). Each point is averaged over $1000$ realizations.
}
\label{hyst3D}
\end{figure}

One of the classical hallmarks of discontinuous phase transitions is
the presence, near the transition, of the hysteresis phenomenon: ramping the
control parameter at a fixed (slow) rate up and down through 
the transition point, a hysteresis loop is formed, inside which phase coexistence
is manifest (see Fig.~\ref{hyst3D}a for the $d=3$ case with vectorial noise). 
The size of such hysteresis loops
varies with the ramping rate. An intrinsic way of assessing phase 
coexistence and hysteresis is to study systematically the
nucleation time $\tau_{\uparrow}$ needed to jump from 
the disordered phase to the ordered
one, as well as $\tau_{\downarrow}$ the decay time after 
which the ordered phase falls into the disordered one. 
Fig.~\ref{hyst3D}b shows, in three space dimensions, 
how these nucleation and decay vary with $\eta$ at two different sizes. 
A sharp divergence is observed, corresponding to the transition point.
At a given time value $\tau$, one can read, from the distance between the 
``up'' and the ``down'' curve, the average size of hysteresis loops for 
ramping rates of the order of $1/\tau$. 

\subsection{Phase diagram}

The above detailed FSS study would be very tedious to realize when
varying systematically the main parameters $\eta$, $\rho$, and $v_{\rm
0}$, as well as the nature of the noise and the presence or not of
repulsive interactions. From now on, to characterize the discontinuous
nature of the transition, we rely mainly on the presence, at
large-enough system sizes $L$, of a minimum in the variation of the
Binder cumulant $G$ with $\eta$ (all other parameters being fixed). We
call $L^*$ the crossover size marking the emergence of a minimum of
$G(\eta)$. 

We are now in the position to sketch the phase diagram in the ($\eta$,
$\rho$, $v_{\rm 0}$) parameter space.  The numerical protocol used is,
at given parameter values, to run a large-enough system so that the
discontinuous character of the transition is seen (i.e. $L>L^*$). For
larger sizes, the location of the transition point typically varies
very little, so that for most practical purposes, locating the
(asymptotic) transition point from systems sizes around $L^*$ is
satisfactory.

\begin{figure}
\includegraphics[width=8.6cm,clip]{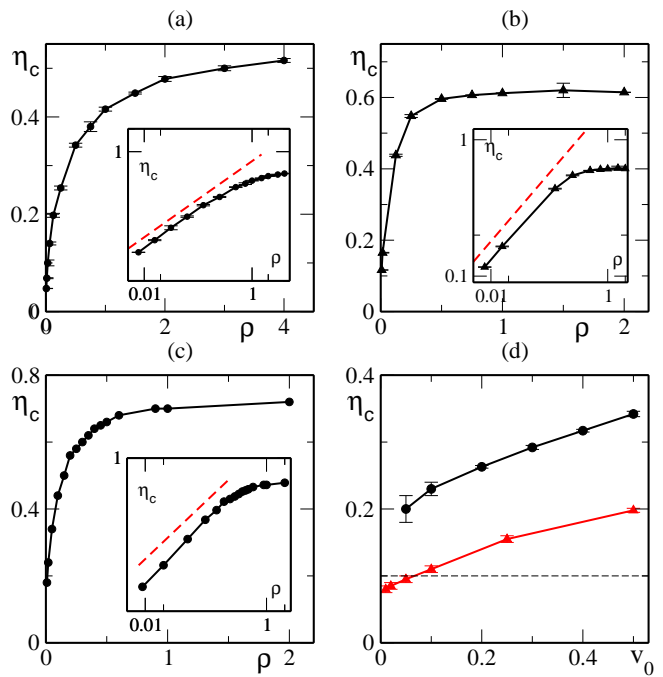}
\caption{\label{Phase}
(color online)
Asymptotic phase diagrams for the transition to collective motion.
(a) Two space dimensions: threshold amplitude $\eta_{\rm t}$ for angular noise as 
a function of density $\rho$ at $v_{\rm 0}=0.5$. 
Inset: Log-log plot to compare the low density behavior with the 
mean field predicted behavior $\eta_{\rm t} \sim \sqrt{\rho}$ (dashed red line).
(b) As in panel (a), buth with vectorial noise dynamics.
(c) Noise-density phase diagram in three dimensions for vectorial
noise dynamics at fixed velocity $v_{\rm 0}=0.5$.
In the log-log inset the transition line can be compared with the predicted
behavior $\eta_{\rm t} \sim \rho^{1/3}$ (dashed red line).
(d) Two space dimensions: threshold amplitude $\eta_{\rm t}$ 
for angular noise as a function of particle 
velocity $v_{\rm 0}$ at fixed density $\rho=1/2$ (black circles) and
$\rho=1/8$ (red triangles). The horizontal dashed line marks the noise amplitude considered 
in Ref. \cite{Nagy} (see Section \ref{SumTransition}).} 
\end{figure}

\begin{figure}
\includegraphics[width=8.6cm,clip]{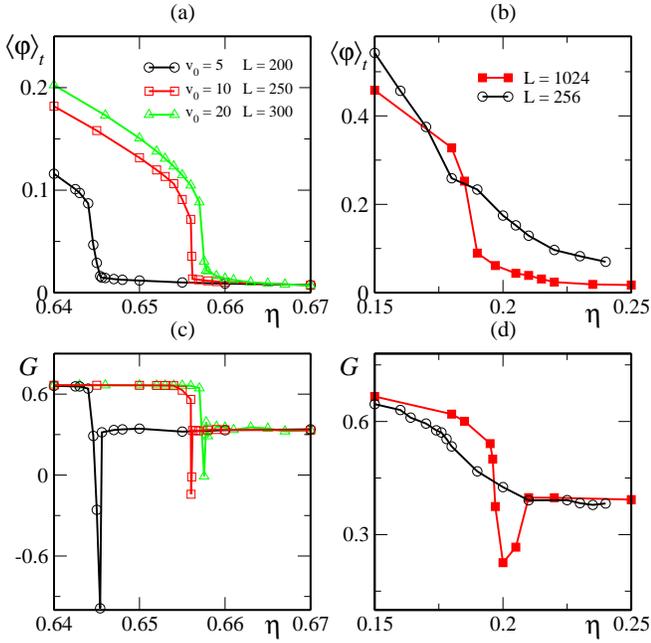}
\caption{\label{SpecialCases}
(color online)
First-order transition for angular noise dynamics at high (left
panels) and low velocity $v_{\rm 0}$ (right panels). Typical averaging time
is $\approx 10^6$ timesteps. (a) Time-averaged order parameter  and
(c) Binder cumulant at large particle velocity for angular noise in
two spatial dimensions at increasing velocities and $L\gtrsim L*$
($\rho=2$). (b) Time averaged order parameter and (d) Binder cumulant
for $v_{\rm 0}=0.05$ and two increasing system sizes ($\rho=1/2$).}
\end{figure}

The results presented below are in agreement with simple
mean-field-like arguments in the diluted limit: in the small-$\rho$
regime, one typically expects that the lower the density, the lower
the transitional noise amplitude $\eta_{\rm t}$.  Indeed, for $\Delta
t \,v_{\rm 0}$ of the order of or not much smaller than the
interaction range $r_{\rm 0}$ and in the low-density limit $\rho \ll
1/{r_{\rm 0}}^d$, the system can be seen as a dilute gas in which
particles interact by short range ordering forces only. In this
regime, the persistence length of an isolated particle (i.e. the
distance travelled before its velocity loses correlation with its
initial direction of motion) varies like $v_{\rm 0}/\eta$.  To allow
for an ordered state, the noise amplitude should be small enough so
that the persistence length remains larger than the average
inter-particle distance, {\it i.e.} $1/\rho^{1/d}$. Thus the
transition noise amplitude is expected to behave as
\begin{equation}
\label{meanfield}
\eta_{\rm t} \, \sim \, v_{\rm 0}\, \rho^\frac{1}{d} \;. 
\end{equation}

In~\cite{Czirok1997}, it was indeed found that
$\eta_{\rm t}\sim\rho^{\alpha}$ with $\alpha\simeq\frac{1}{2}$ in two dimensions.
Our own data (Fig. \ref{Phase}a-c) now confirm Eq. (\ref{meanfield}) 
for both the angular and vectorial
noise in two and three spatial dimensions, 
down to very small $\rho$ values. 
The data deviate from the square-root behavior as the average
inter-particle distance gets of the order of or 
smaller than the interaction range.  

Finally, we also investigated the transition line when $v_{\rm 0}$ is
varied (Fig. \ref{Phase}d). For the vectorial noise case, at fixed
density, the threshold noise value $\eta_{\rm t}$ is almost constant
(data obtained at $\rho=\frac{1}{2}$, not shown).  For the angular
noise, in the small $v_{\rm 0}$ limit where the above mean-field
argument does not apply, we confirm the first-order character of the
phase transition down to $v_{\rm 0} \approx 0.05$ for both angular and
vectorial noise (Fig.~\ref{SpecialCases}b,d).  For even smaller values
of $v_{\rm 0}$, the investigation becomes numerically too costly (see
section below).  Note that $\eta_{\rm t}$ seems to be finite when
$v_{\rm 0}\to 0^+$, a limit corresponding to the XY model on a
randomly connected graph.  Still for angular noise, the large velocity
limit is also difficult to study numerically. Again, we observe that
the transition is discontinuous as far as we can probe it, i.e.
$v_{\rm 0}=20$ (Fig.~\ref{SpecialCases}a,b).

\subsection{Special limits and strength of finite-size effects}

We now discuss particular limits of the models above together with the
relative importance of finite-size effects. Recall that these are quantified
by the estimated value of the crossover size $L^*$ beyond which the transition 
appears discontinuous. All the following results have been obtained
for $d=2$. Partial results in three dimensions indicate that the same
conclusions should hold there. Keep in mind that in all cases reported, 
the transition is discontinuous. We are just interested here in how 
large a system one should use in order to reach the asymptotic regime.

\begin{figure}
\includegraphics[width=8.6cm,clip]{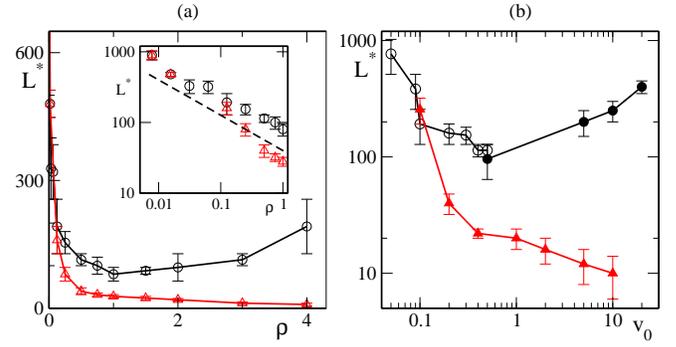}
\caption{\label{FS}
(color online) Crossover system size $L^*$ above which 
the discontinuous character of the transition appears (as testified by the existence
of a minimum in the $G(\eta)$ curve). Black circles: angular
noise. Red triangles: vectorial noise.
(a) $L^*$ vs $\rho$ for $v_{\rm 0}=0.5$;
Inset: the low density behavior in log-log scales;
the dashed line marks a power-law divergence proportional to $1/\sqrt{\rho}$.
(b) $L^*$ vs $v_{\rm 0}$ at fixed density (open symbols: $\rho = 1/2$;
filled symbols $\rho = 2.0$).}
\end{figure}

Fig.~\ref{FS}a shows that finite-size effects are 
stronger for angular noise than for vectorial noise for all densities $\rho$
at which we are able to perform these measurements.
Note in particular that at $\rho=2$, the density originally used by 
Vicsek {\it et al.}, $L^*\sim 128$ for angular noise, 
while it is very small for 
vectorial noise, confirming the observation made in Section \ref{overture}.

In the small-$\rho$ limit, the discontinuous 
character of the transition appears later and later, 
with $L^*$ roughly diverging as $1/\sqrt{\rho}$ (inset of Fig.~\ref{FS}a).  
Note that this means that in the small-$\rho$ limit
one needs approximately the same number of particles to start observing 
the discontinuity.

The large-$\rho$ limit reveals a difference between angular and vectorial
noise: while $L^*$ remains small for vectorial noise, it seems to diverge
for angular noise (Fig.~\ref{FS}a), making this case difficult to study numerically.

We also explored the role the microscopic velocity $v_{\rm 0}$ in the strength
of finite-size effects. 
Qualitatively, the effects observed are similar to
those just reported when the density is varied  (Fig.~\ref{FS}b). 
In the small $v_{\rm 0}$ limit, we record a strong increase of $L^*$ as $v_{\rm 0}\to 0$
for both types of noise. In the large velocity limit, $L^*$ decreases for 
vectorial noise, whereas it increases for angular noise. 

\subsection{Summary and discussion}
\label{SumTransition}

The summary of the above lengthy study of the order-disorder transition 
in Vicsek-like models is simple: for any finite density $\rho$, any 
finite velocity $v_{\rm 0}$, and for both types of noise introduced, 
the transition is discontinuous. This was observed 
even in the numerically-difficult limits of large or small
$\rho$ or $v_{\rm 0}$.
These results contradict recent claims made about the angular noise case
(original Vicsek model). We now comment on these claims.

Vicsek and co-workers~\cite{Nagy}, showed that keeping the
density and the noise intensity fixed, a qualitative change is
observed when decreasing $v_{\rm 0}$: for not too small $v_{\rm 0}$ values, in the
ordered phase, particles diffuse anisotropically (and the transition
is discontinuous), while diffusion becomes isotropic at small $v_{\rm 0}$,
something interpreted as a sign of a continuous transition in this
region. Rather than the convoluted arguments presented there, what
happens is in fact rather simple: decreasing $v_{\rm 0}$ at fixed $\rho$ and
$\eta$, one can in fact cross the transition line, passing from the
ordered phase (where particles obviously diffuse anisotropically
due to the transverse superdiffusive effects discussed in Section \ref{superdiff}) to
the disordered phase. Our Figure~\ref{Phase}d, obtained in the same
conditions as in \cite{Nagy} (apart from harmless change of time
updating rule), shows that if one keeps $\eta=0.1$ (as in
\cite{Nagy}), one crosses the transition line at about $v_{\rm 0}\simeq
0.1$, the value invoked by Vicsek and co-workers to mark a crossover from
discontinuous to ``continuous'' transitions.

In a recent letter~\cite{Aldana2007}, Aldana {\it et al.} study
order-disorder phase transitions in random network models and show
that the nature of these transitions may change with the way noise is
implemented in the dynamics (they consider the angular and vectorial
noises defined here).  Arguing that these networks are limiting cases
of Vicsek-like models, they claim that the conclusions reached for the
networks carry over to the transition to collective motion of the
VM-like systems.  They conclude in particular that in the case of
``angular'' noise the transition to collective motion is continuous.
We agree with the analysis of the network models, but the claim that
they are relevant as limits of Vicsek-like models is just wrong: the
data presented there (Figure~1 of~\cite{Aldana2007}) to substantiate
this claim is contradicted by our Figures~\ref{SpecialCases}a,c (see
also~\cite{Chate2007}) obtained at larger system sizes.  
Again, for large-enough system sizes, the
transition is indeed discontinuous. Thus, at best, the network models
of Aldana {\it et al.}  constitute a singular $v_{\rm 0}\to\infty$ limit of
Vicsek-like models.

\section{Nature of the ordered phase}
\label{ordered}

We now turn our attention to the ordered, symmetry-broken phase.  In
previous analytical studies, it has often been assumed that the
density in the ordered phase is spatially homogeneous, albeit with
possibly large fluctuations (see, e.g.~\cite{Toner1995}).  This is
indeed what has been reported in early numerical studies, in
particular by Vicsek {\it et al.}~\cite{Vicsek1995}. In the following, we show
that this is not true in large enough systems, where, for a wide range
of noise amplitudes near the transition point, density fluctuations
lead to the formation of localized, travelling, high-density and
high-order structures.  At low enough noise strength, though, a
spatially-homogeneous ordered phase is found, albeit with unusually
strong density fluctuations.

\subsection{Traveling in bands}

\begin{figure}
\includegraphics[width=8cm,clip]{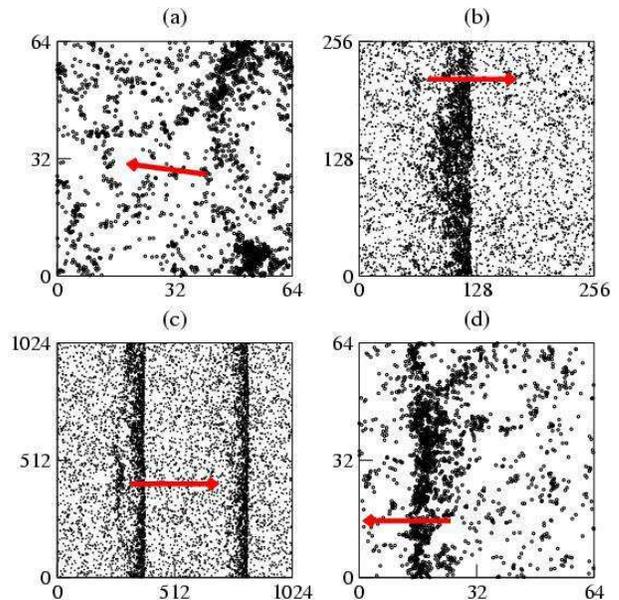}
\caption{(color online)
Typical snapshots in the ordered phase. 
Points represent the position of individual particles and the red arrow points
along the global direction of motion. (a)-(c): Angular noise, $\rho=1/2$,
$v_{\rm 0}=0.5$, $\eta = 0.3$ and increasing system 
sizes - respectively $L=64$, $L=256$ and $L=1024$. 
Sharp bands can only be observed if $L$ is larger than the typical band width $w$.
(d) Vectorial noise: $\rho=1/2$, $v_{\rm 0}=0.5$, $\eta = 0.55$ and $L=64$: bands appear
at relatively small system sizes for this type of noise. 
For clarity reasons, only a representative sample of 
10000 particles is shown in (b) and (c). Boundary conditions are periodic.}
\label{Photo1}
\end{figure}

\begin{figure}
\includegraphics[width=8.6cm,clip]{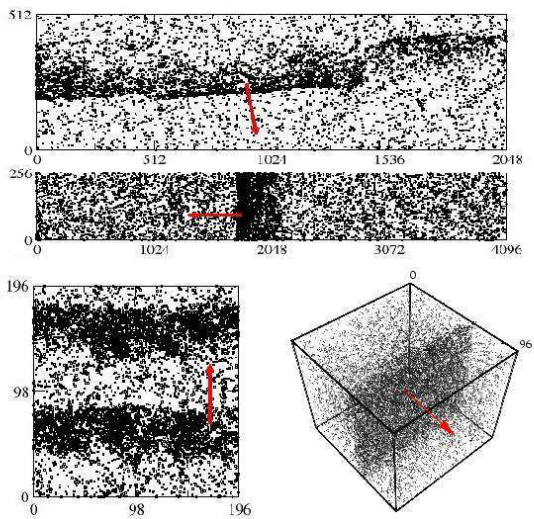}
\caption{(color online)
Same as Fig.~\ref{Photo1} but in different geometries/boundary conditions or
space dimensions.
(a)-(b) vectorial noise ($\eta = 0.325$, $\rho=1/8$, and $v_{\rm 0}=0.5$),
boundary conditions are periodic along the $y$ (vertical) axis
and reflecting in $x$. (a): a long single band travels along the 
periodic direction.
(b): the domain size along the periodic direction is too small to accomodate bands, 
and a single band bouncing back and forth along the non-periodic direction is observed.
(c) Angular noise, repulsive force, and periodic boundary conditions
($\rho=2$, $\eta=0.23$ and $v_{\rm 0}=0.3$).
(d) High-density sheet traveling in a three-dimensional box with periodic 
boundary conditions (angular noise with amplitude 
$\eta=0.355$, $\rho=1/2$ and $v_{\rm 0}=0.5$).}
\label{Photo2}
\end{figure}

Numerical simulations of the ordered phase dynamics ($\eta < \eta_{\rm t}$), 
performed at large enough noise amplitudes, are characterized 
by the emergence of high-density moving bands ($d=2$) or sheets
($d=3$). Typical examples are given in Figs.~\ref{Photo1}--\ref{Photo2}.
These moving structures appear for large-enough systems after some transient.
They extend transversally with respect to the mean direction of motion, 
and have a center of mass velocity close to $v_{\rm 0}$.
While particles inside bands are ordered and, in the asymptotic regime, move coherently 
with the 
global mean velocity, particles lying outside bands ---in low density regions--- 
are not ordered and perform random walks.

As shown in Fig.~\ref{Photo1}a-c for angular noise dynamics (\ref{EVicsek}), 
there exists a typical system size $L_{\rm b}$, 
below which the bands or sheets cannot be observed.   
Numerical simulations indicate that $L_{\rm b}$ depends only weakly on the
noise amplitude and is of the same order of magnitude as the 
crossover size marking the appearance of the discontinuous character of the transition:
$L_{\rm b} \approx L^*$.
It is therefore numerically easier to observe bands in the ordered phase of vectorial 
noise dynamics (\ref{EN2}), as in Fig~\ref{Photo1}d.

Bands may be observed asymptotically
without and with a repulsive interaction (Fig~\ref{Photo2}c) and for both kinds of noise.
They appear for various choices of boundary conditions
(see for instance Fig. \ref{Photo2}a-b, where reflecting boundary conditions
have been employed), which may play a role in determining 
the symmetry-broken mean direction  
of motion. For instance, bands travelling parallel to one of the axis are favoured
when periodic boundary conditions are employed 
in a rectangular box (they represent the simplest way in which an extended 
structure can wrap around a torus, and are thus reached more easily from 
disordered initial conditions), 
but bands travelling in other directions may also appear, albeit with a 
smaller probability.

\begin{figure}
\includegraphics[width=8cm,clip]{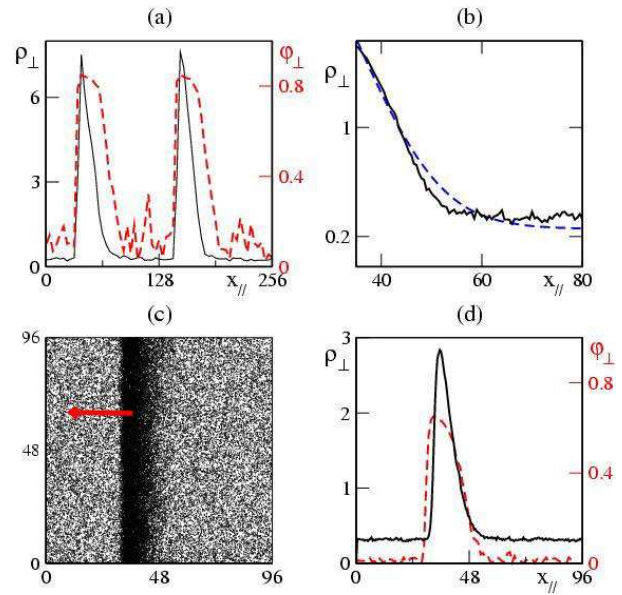}
\caption{(color online)
(a) Typical density (black line) and order parameter (dashed red line) profiles for
bands in two dimensions (vectorial noise, $\rho = 2$, $\eta = 0.6$ and $v_{\rm 0} = 0.5$).
(b) Tail of the density profile shown in (d) (black line) and its fit
(blue dashed line) by the formula: 
$\rho_{\perp}(x_{\parallel}, t) \approx a_0 + a_1(t)\,\exp(-x_{\parallel}/w)$,
with $w \approx 6.3$ (lin-log scales).
(c-d) Traveling sheet in three dimensions (angular noise, 
$\rho = 1/2$, $\eta = 0.355$, and $v_{\rm 0} = 0.5$).
(c) projection of particle positions on a plane containing the global
direction of motion (marked by red arrow).
(d) Density (black line) and order parameter (dashed red line) profiles along 
the direction of motion $x_{\parallel}$.}
\label{Fband}
\end{figure}

Bands can be described quantitatively through local quantities, 
such as the local density
$\rho_{\ell}(\vec{x}, t)$, measured inside a domain
$\mathcal{V}(\vec{x})$ centered around $\vec{x}$, and the local order parameter
\begin{equation}
\varphi_{\ell}(\vec{x}, t)
=\frac{1}{v_{\rm 0}}\left|\langle\vec{v_i}(t)\rangle_{\vec{r}_i \in \mathcal{V}(\vec{x})}\right|\,.
\end{equation}
Further averaging these local quantities perpendicularly to the mean velocity
(\ref{globalop}) one has the  
{\it density profile} 
$\rho_{\perp}(x_{\parallel}, t) = \langle \rho_{\ell}(\vec{x},t) \rangle_{\perp}$
and the
{\it order parameter profile}  
$\varphi_{\perp}(x_{\parallel},t) = \langle \varphi_{\ell}(\vec{x},t) \rangle_{\perp}$,
where $x_{\parallel}$ indicates the longitudinal direction w.r.t. mean velocity.
Bands are characterized by a sharp kink in both the density and the order parameter
profiles (see Fig. \ref{Fband}a and \ref{Fband}c-d). 
They are typically asymmetric, as it can be expected for
moving structures, with a rather sharp front edge, a well defined
mid-height width $w$ -- which typically is of the same order as $L_b$ -- 
and an exponentially decaying tail with a characteristic
decay length of the order of the $w$ (Fig.~\ref{Fband}b).

\begin{figure}[!ht]
\includegraphics[width=8.6cm,clip]{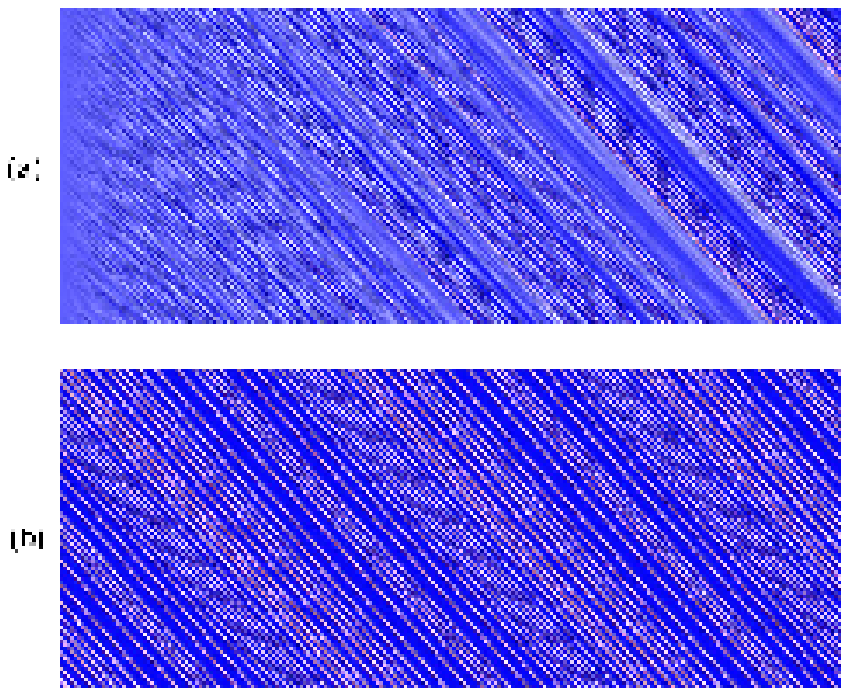}
\includegraphics[width=8.6cm,clip]{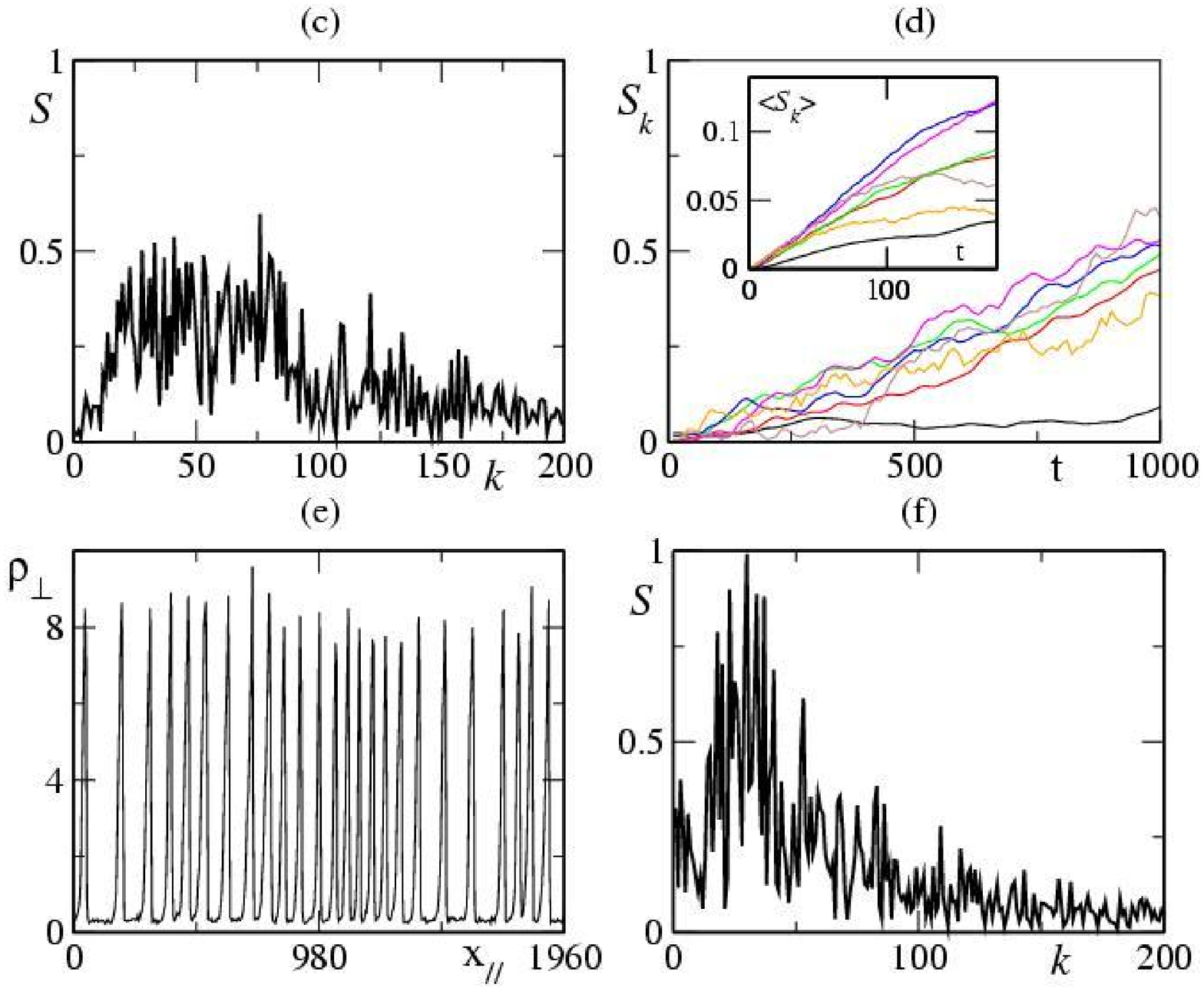}
\caption{(color online)
Emergence of high-density high-order traveling bands ($d=2$) from a 
spatially-homogeneous (uniformely distributed random positions) initial condition with all particle
velocities oriented along the major axis of a $196\times 1960$ domain with periodic
boundary conditions. Vectorial noise of amplitude $\eta = 0.6$, density $\rho=2$,
and $v_{\rm 0}=0.5$.  
(a): space-time plot of the density profile 
(time is running from left to right from $t=0$ to $t=12000$, while the longitudinal direction is represented in
ordinates. Colour scale from blue (low values)
to red (high values).).
(b): same as (a) but at later times (from $t=148000$ to $t=160000$).
(c): spatial Fourier power spectrum $S$ of an early density profile ($t=12000$).
(d): early time evolution of selected Fourier modes (chosen between spectrum peaks of (c)), 
$k = 10, 23, 28, 33, 41, 76, 121$ (inset: average over 50 different runs). 
(e): density profile at a late time ($t=160000$, final configuration of (b)).
(f): same as (c) but for the late time density profile of (e). The quality
of this figure has been reduced to meet the ArXiv size constraint.
}
\label{LongBand}
\end{figure}

Large systems may accomodate several bands at the same time, typically 
all moving in the same direction (see for instance Fig. \ref{Photo1}c 
and the density profile in Fig.~\ref{LongBand}e). 
However, they do not form well-defined wave trains, 
but rather a collection of solitary objects, as hinted by the following numerical
experiments.

We investigated the instability of the density-homogeneous, ordered
state in a series of numerical simulations starting from particles
uniformly distributed in space but strictly oriented along the major axis in
a large rectangular domain.  Figure \ref{LongBand}a,b show space-time
plots of the density profile: initially flat, it develops structures
with no well-defined wavelength (Fig.~\ref{LongBand}c).  Density
fluctuations destroy the initially-ordered state in a rather unusual
way: a dynamical Fourier analysis of the density profile show a
weakly-peaked, wide band of wavelengths growing {\it subexponentially}
(Fig. \ref{LongBand}d). This is at odds with a finite-wavelength
supercritical instability, which would lead to a wavetrain of
traveling bands.  Furthermore, the asymptotic (late time) power
spectra of the density profiles are not peaked around a single
frequency either, but rather broadly distributed over a large range of
wavenumbers (Fig.~\ref{LongBand}f).  In the asymptotic regime, bands
are extremely long-lived metastable (or possibly stable) objects,
which are never equally-spaced (a typical late-time configuration is
shown in Fig. \ref{LongBand}e).

To summarize, the emerging band or sheet structure 
in the asymptotic regime is not a 
regular wave train characterized by a single wavelength, 
but rather a collection of irregularly-spaced localized traveling objects,
probably weakly interacting through their exponentially decaying tails.

\subsection{Low-noise regime and giant density fluctuations}

As the noise amplitude is decreased away from the transition point,
bands are less sharp, and eventually disappear, giving way to an ordered state
characterized by an homogeneous local order parameter 
and large fluctuations of the 
local density.

A quantitative measure of the presence, in the ordered phase, 
of structures spanning the dimension transverse to the mean motion
(i.e. bands or sheets) is provided by the variances  
of the density and order parameter profiles: 
\begin{equation}
\begin{array}{lll}
\label{width}
\Delta\rho^2_{\perp}(t) &=& \langle (\rho_{\perp}(x_{\parallel}, t) 
- \langle \rho_{\perp}(x_{\parallel}, t)\rangle_{\parallel})^2 \rangle_{\parallel}\\
\Delta\varphi^2_{\perp}(t) &=& \langle (\varphi_{\perp}(x_{\parallel}, t) - 
\langle \varphi_{\perp}(x_{\parallel}, t)\rangle_{\parallel})^2 \rangle_{\parallel}
\end{array}
\end{equation}
where $\langle \cdot \rangle_{\parallel}$ indicates the average of the 
profile in the longitudinal direction with respect to mean velocity. 
Indeed, these profile widths vanish in the infinite-size limit except if band/sheet
structures are present.

\begin{figure}
\includegraphics[width=8.6cm,clip]{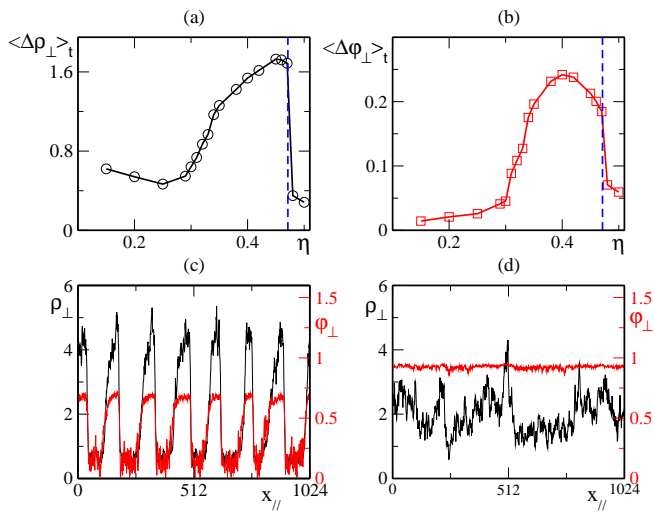}
\caption{(color online)
(a-b) Time-averaged profile width for both density (a) 
and order parameter (b)
as a function of noise amplitude in the ordered phase
(angular noise, $1024\times 256$ domain, global motion along the major axis,
$\rho=2$, and $v_{\rm 0}=0.5$).
The dashed vertical blue line marks the order-disorder transition. 
(c-d) Typical instantaneous profiles along the long dimension of the system
described in (a-b) for intermediate noise value ((b) $\eta=0.4$) and
in the bandless regime ((c) $\eta=0.15$).}
\label{BandGNF}
\end{figure}

In Fig \ref{BandGNF}a,b, we plot these profile widths averaged over time
as a function of noise amplitude.  Both quantities present a maximum
close to the transition point in the ordered phase, and drop
drastically as soon as the disordered phase is entered.  Lowering the
noise away from the transition point, these profiles decrease
steadily: bands/sheets stand less sharply out of the disordered
background (Fig~\ref{BandGNF}c).  At some point ($\eta \approx 0.3$
for the parameters values considered in Fig~\ref{BandGNF}a,b), bands
rather abruptly disappear and are no more well-defined transversal
objects.  It is difficult to define this point accurately, but it is
clear that for lower noise intensities the local order parameter is
strongly homogeneous in space.  Nevertheless, fluctuations in the
density field are strong (Fig~\ref{BandGNF}d), but can no more give
rise to (meta)stable long-lived transverse structures.

\begin{figure}
\includegraphics[width=8.6cm,clip]{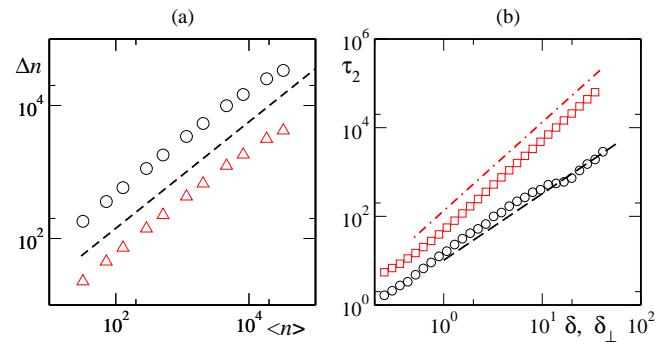}
\caption{(color online) Giant density fluctuation and transverse
superdiffusion in the bandless ordered phase.  (a) anomalous density
fluctuations (see text): $\Delta n$ scales
approximately like $n^{0.8}$ (the dashed line has slope 0.8) both in
two dimensions (black circles, $L=256$, $\rho=2$, $v_{\rm 0}=0.5$,
angular noise amplitude $\eta=0.25$) and in three dimensions (red triangles,
$L=64$, $\rho=1/2$ $v_{\rm 0}=0.5$, vectorial noise amplitude $\eta=0.1$, values
shifted for clarity).  (b) average doubling time $\tau_2$
of the transverse (w.r.t. mean velocity) interparticle distance
$\delta_{\perp}$. Black circles: ordered bandless regime ($\rho=4$, angular
noise amplitude $\eta=0.2$ in a rectangular box of size $1024 \times
256$). The black dashed line marks the expected growth $\tau_2 \sim
\delta_{\perp}^{3/2}$.  Red squares: same but deep in the disordered phase
($\rho=4$, angular noise
amplitude $\eta=1$, $L=512$). The dot dashed red line shows normal
diffusive behavior: $\tau_2 \sim \delta^2$.}
\label{SDF-GNF}
\end{figure}

Density fluctuations in the bandless regime are in fact anomalously
strong: measuring number fluctuations in sub-systems of linear size
$\ell$, we find that their root mean-square $\Delta n$ does {\it not}
scale like the square root of $n=\rho\ell^d$, the mean number of
particles they contain; rather we find $\sigma(n) \propto n^\alpha$
with $\alpha\approx 0.8$ both in $d=2$ and in $d=3$
(Fig.~\ref{SDF-GNF}a).  This is reminiscent of the recent discovery of
``giant density fluctuations'' in active nematics~\cite{Mishra,
Narayan, Chate2006}. However, the theoretical argument which initially
predicted such fluctuations~\cite{Ramaswamy} cannot be invoked
directly in the present case. (Indeed, the above value of $\alpha$,
although needing to be refined, does not seem to be compatible with
the prediction $\alpha=\frac{1}{2}+\frac{1}{d}$ made
in~\cite{Ramaswamy}.)  More work is needed to fully understand under
which circumstances the coupling between density and order in systems
of ``active'', self-propelled particles gives rise to such anomalous
density fluctuations.

\subsection{Transverse superdiffusion}
\label{superdiff}

According to Toner and Tu predictions \cite{Toner1995,Toner1998,Toner1998_2},
the dynamics of the symmetry-broken ordered phase 
of polar active particles should be characterized by a superdiffusive mean
square displacement 
\begin{equation}
\Delta x_{\perp}=\sqrt{\langle (x_{\perp}(t)-x_{\perp}(0))^2\rangle_i} 
\end{equation} 
in the direction(s) transversal to mean velocity. In particular, in $d=2$ one has
~\cite{Toner1998_2} 
\begin{equation}
\label{superd}
\Delta x_{\perp}^2 \sim t ^{\nu}
\end{equation}
with $\nu=4/3$.  While this analytical result has been succesfully
tested by numerical simulations of models with cohesive interactions
\cite{Toner1998_2, Gregoire2004}, numerical simulations in models
without cohesion present substantial difficulties, mainly due to the
presence of continuously merging and splitting
sub-clusters of particles moving coherently (as discussed in Section
\ref{cluster}). As a consequence, an ensemble of test particles in a
cohesionless model is exposed to different ``transport'' regimes
(w.r.t. center of mass motion) which are not well separated
in time. When the mean displacement is averaged at fixed time, this tends
to mask the transverse superdiffusion.  

To overcome this problem, we
chose to follow \cite{Vulpiani2000} and to measure $\tau_2$, the
average time taken by two particles to double their transverse
separation distance $\delta_{\perp}$.  From Eq. (\ref{superd}) one
immediately has
\begin{equation}
\label{superd2}
\tau_2 \sim \delta_{\perp}^{2/\nu} 
\end{equation} 
with $2/\nu=3/2$ in $d=2$.
In order to easily separate the transverse from the
parallel component, we considered an ordered system in a large
rectangular domain with periodic boundary conditions and the mean
velocity initially oriented along the long side.  The mean direction
of motion then stays oriented along this major axis, 
so that we can identify the transverse direction with the
minor axis.  Furthermore, a high density and a small (angular) noise
amplitude (corresponding to the bandless regime) have been chosen to
avoid the appearance of large, locally disordered patches. \\ Our
results (Fig.~\ref{SDF-GNF}b) confirm the prediction of Toner and Tu:
transverse superdiffusion holds at low-enough noise, while normal
diffusion is observed in the disordered, high-noise phase.  Note that
the systematic deviation appearing in our data at some large scale is
induced by large fluctuations in the orientation of the global mean
velocity during our numerical simulations (not shown here).

We take the opportunity of this discussion to come back to the 
superdiffusive behavior of particles observed in the 
transition region\cite{Comment_gg}.
There, sub-clusters emerge propagate ballistically and
isotropically due to the absence of a well-established global
order. Particles trajectories consist in ``ballistic flights'',
occurring when a particle is caught in one of these 
coherently moving clusters,
alternated with ordinary diffusion in disordered regions. The mean
square displacement of particles exhibits the 
scaling $\Delta x^2 = \langle |\vec{x}(t)-\vec{x}(0))|^2\rangle_i
\propto t ^{5/3}$ \cite{Comment_gg}. In view of our current understanding
of the discontinuous nature of the transition, we now tend to believe that
this isotropic superdiffusion is probably {\it not} asymptotic.

\subsection{Internal structure of the ordered region}
\label{cluster}

We now turn our attention to the internal structure of the ordered regimes. 
As we noted in the previous section, these regimes do {\it not}
consist of a single cluster of interacting particles moving coherently. 
Even in the case where 
high density bands/sheets are present, these are in fact dynamical objects made of 
splitting and merging clusters. Note that for the models considered here, clusters
are unambiguously defined thanks to the strictly finite interaction range $r_{\rm 0}$.

\begin{figure}
\includegraphics[width=8.6cm,clip]{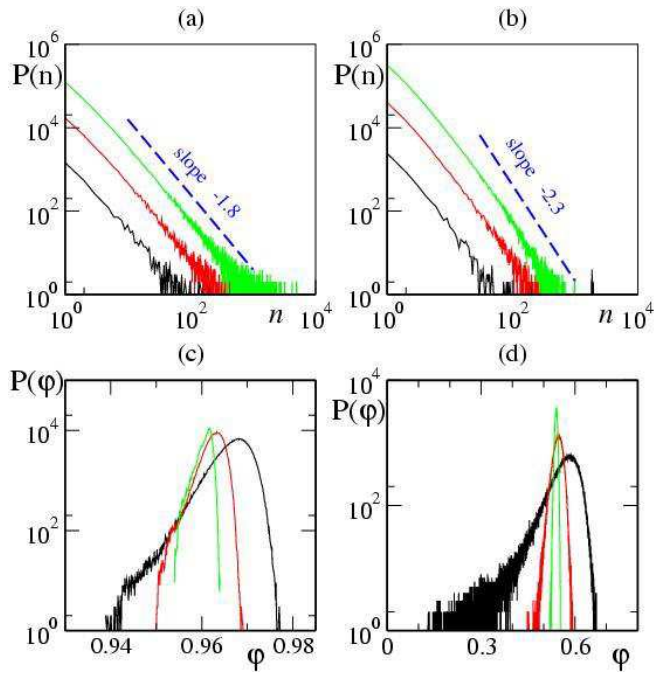}
\caption{
(a-b) Cluster size distributions (arbitrary units) for domain sizes 
$L=32$ (black), 128 (red) , and 512 (green) from left to right
($d=2$, $\rho=2$, $v_{\rm 0}=0.5$, angular noise). 
(c-d) Probability distribution functions of the order parameter $\varphi$ 
(arbitrary units) for the same parameters and system sizes as in (a-b).
(The most peaked distributions are for the largest size $L=512$.)
Left panels (a,c): $\eta=0.1$, bandless regime; Right panels (b,d), 
regime with bands at $\eta=0.4$.}
\label{FigCluster}
\end{figure}

As noticed first by Aldana and Huepe~\cite{Huepe}, clusters of size $n$ are distributed
algebraically in the ordered region, i.e. $P(n) \sim n^{-\mu}$. But a closer look reveals that the exponent 
$\mu$ characterizing the distribution of cluster sizes changes with the distance to 
the transition point. For noise intensities not too far from the threshold, when bands
are observed, we find $\mu$ values larger than 2, whereas $\mu<2$ in the bandless 
regimes present at low noise intensities (Fig.~\ref{FigCluster}a,b). 

Thus, bands are truly complex, non-trivial
structures emerging out of the transverse 
dynamics of clusters with a well defined
mean size (since $\mu>2$). It is only in the bandless regime that one can speak,
as Aldana and Huepe, of ``strong intermittency''.
We note in passing that the parameter values they considered
correspond in fact to a case where bands are easily observed 
(at larger sizes than those considered in~\cite{Huepe}). Thus, clusters do have 
a well-defined mean size in their case. Consequently,
the probability distribution $P(\varphi)$ 
of the order parameter $\varphi$ does {\it not} show
the behavior reported in Fig.~1 of~\cite{Huepe} as soon as the system size is 
large enough. Whether in the band/sheet regime or not,
$P(\varphi)$ shows essentially Gaussian tails, 
is strongly peaked around its mean, 
and its variance decreases with system size  
(Fig.~\ref{FigCluster}c,d). 

Although the picture of intermittent bursts between ``laminar''intervals 
proposed by Aldana and Huepe has thus to
be abandoned, the anomalous density fluctuations reported in the
previous section are probably tantamount to the strong intermittency
of cluster dynamics in the bandless regime. Again, these phenomena,
reported also in the context of ``active''
nematics~\cite{Chate2006,Mishra,Ramaswamy}, deserve further
investigation.

\subsection{Phase ordering}
\label{coarsening}

\begin{figure*}[!ht]
\includegraphics[width=5.6cm,clip]{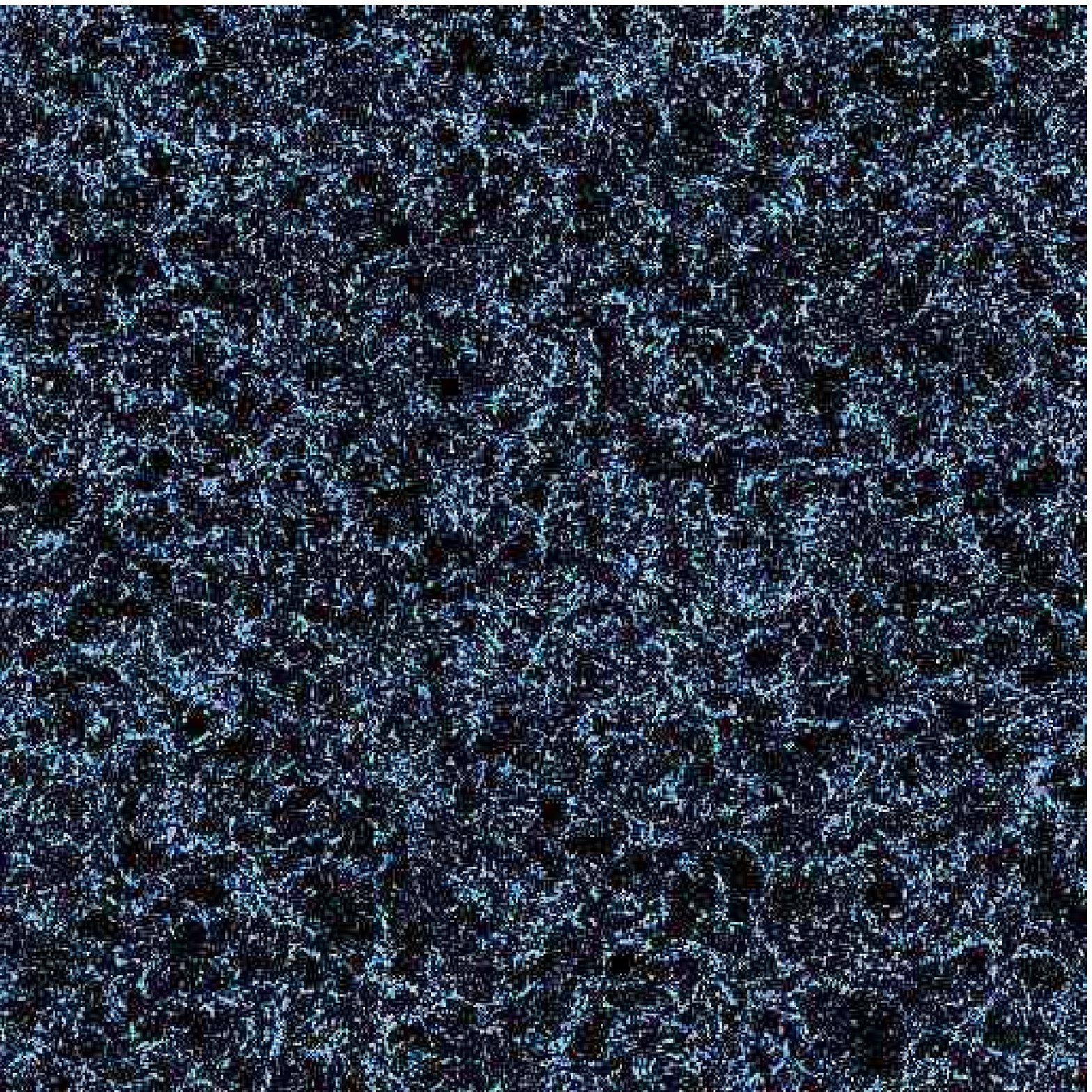}
\includegraphics[width=5.6cm,clip]{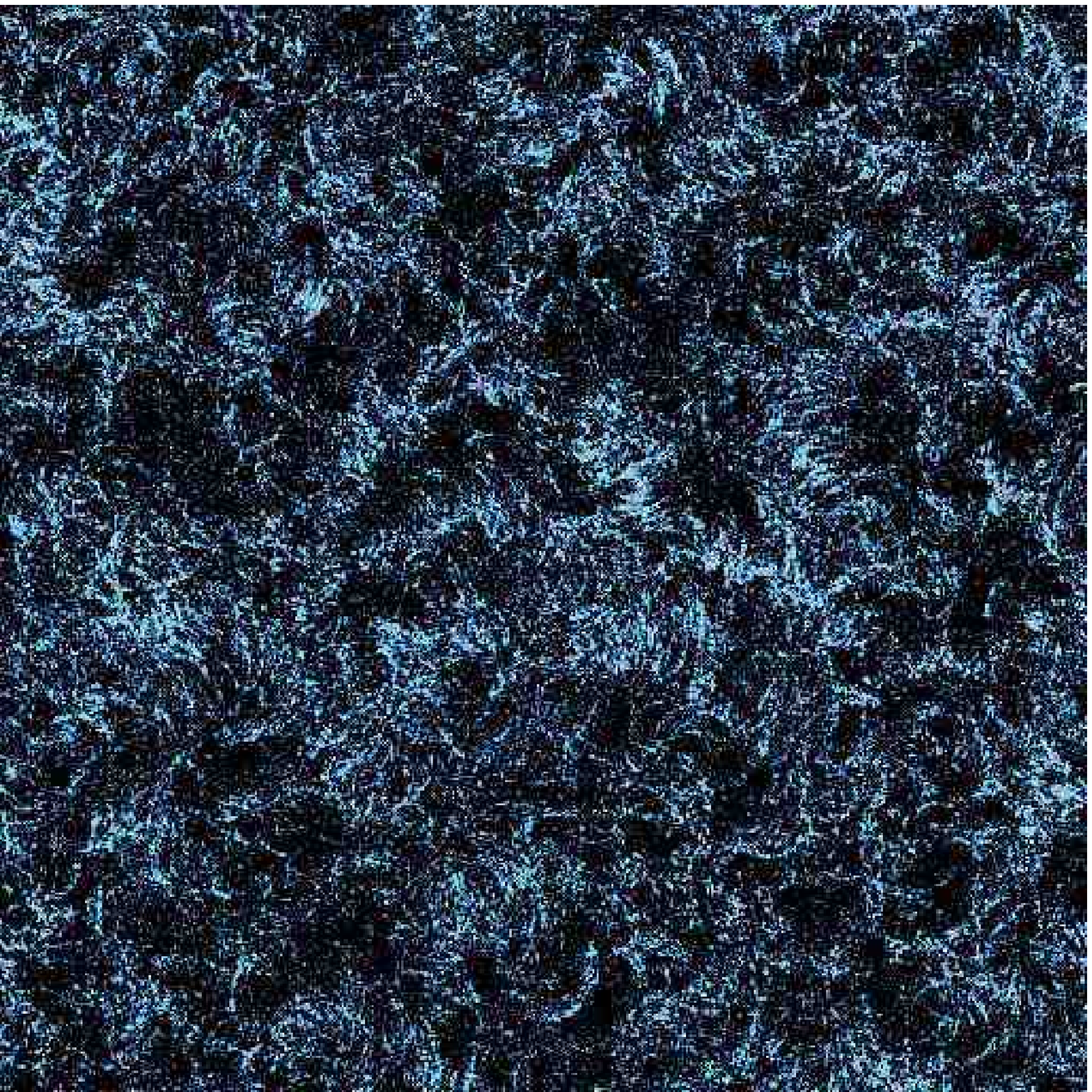}
\includegraphics[width=5.6cm,clip]{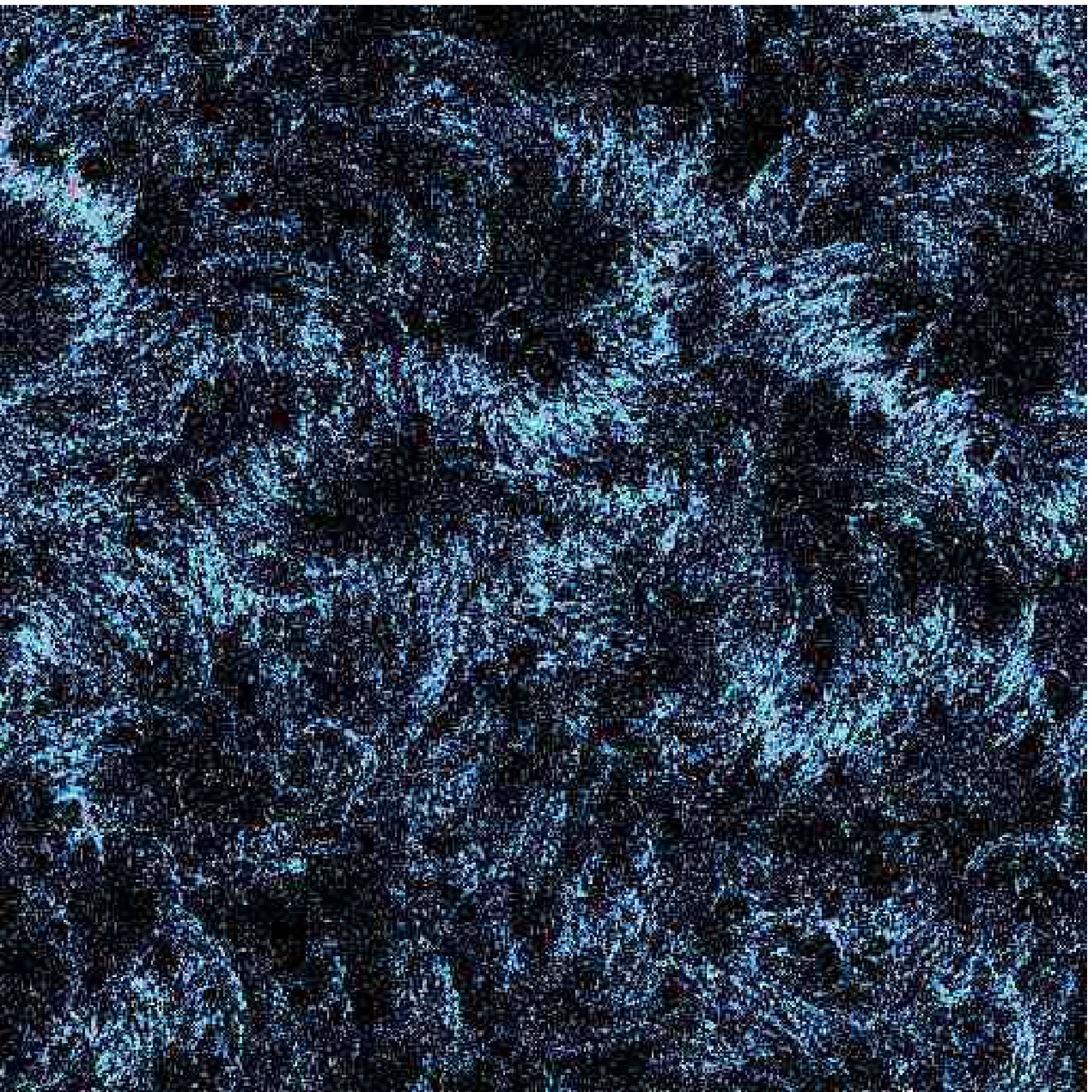}
\caption{(color online) Phase-ordering from disordered initial conditions 
($d=2$, angular noise amplitude $\eta=0.08$, $\rho=1/8$, $v_{\rm 0}=0.5$, system size $L=4096$).
Snapshots of the density field coarse-grained on a scale $\ell=8$ at times $t=160$, 320, and
640 from left to right.}
\label{Coarse-snaps}
\end{figure*}

\begin{figure}
\includegraphics[width=8.6cm,clip]{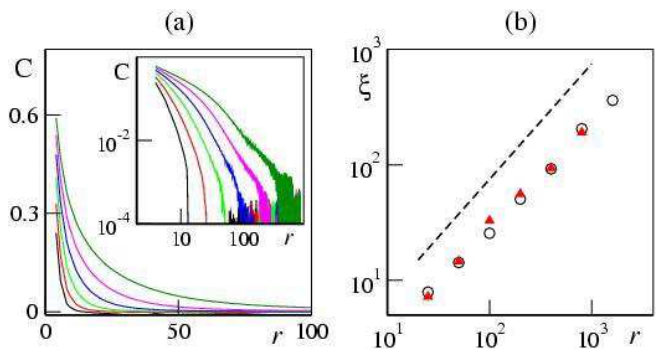}
\caption{(color online) Phase ordering as in Fig.~\ref{Coarse-snaps} ($L=4096$
$\rho=1/2$, $v_{\rm 0}=0.5$). 
(a) two-point
density correlation function 
$C(r,t)=\langle \rho_{\ell}(\vec{x},t)  \rho_{\ell}(\vec{x}+\vec{r},t)\rangle_{\vec{x}}$ 
(coarse grained over a scale $\ell = 4$) as a function of distance $r=|\vec{r}|$ at different time steps:
from left to right 
$t=50,100,200,400,800,1600$. Noise amplitude is $\eta=0.25$, data
have been further averaged over $\approx 40$ different realizations. 
Inset: log scales reveal the intermediate near-algebraic decay 
and the quasi-exponential cut-off.
(b) Lengthscale  $\xi$, estimated from the exponential cut-off positions, 
as a function of time. Empty black circles: $\eta=0.25$ as in (a)
(i.e. regime in which bands are observed asymptotically).
Red full triangles: $\eta=0.1$ 
(i.e. in the bandless regime). 
The dashed black line marks linear growth.}
\label{Coarsening}
\end{figure}

The ordered regimes presented above are the result of some transient evolution.
In particular, the bands/sheets are the typical asymptotic structures 
appearing in {\it finite} domains with appropriate boundary conditions. 
In an infinite system, the phase ordering process is, 
on the other hand, infinite, and worth studying for its own sake.

Numerically, we have chosen to start from highly disordered initial
conditions which have a homogeneous density and vanishing local order
parameter.  In practice, we quench a system ``thermalized'' at strong
noise to a smaller, subcritical, $\eta$ value. Typical snapshots show
the emergence of structures whose typical scale seems to increase fast
(Fig.~\ref{Coarse-snaps}).  During this domain growth, we monitor the
two-point spatial correlation function of both the density and
velocity fields. These fields are defined by a coarse-graining over a
small lengthscale $\ell$ (typically 4).  These correlation functions
have an unusual shape (Fig.~\ref{Coarsening}a): after some rather fast
initial decay, they display an algebraic behavior whose effective
exponent decreases with time, and finally display a near-exponential
cut-off. As a result, they cannot be easily
collapsed on a single curve using a simple, unique, rescaling
lengthscale. Nevertheless, using the late exponential cut-off, a
correlation length $\xi$ can be extracted.  Such a lengthscale $\xi$ grows
roughly linearly with time (Fig.~\ref{Coarsening}b).  Qualitatively
similar results are obtained whether the noise strength is in the
range where bands/sheets appear in finite boxes or not.

We note that the above growth law is reminiscent of that of the
so-called model~H of the classification of Halperin and
Hohenberg~\cite{Hohenberg}. Since this model describes, in principle,
the phase separation in a viscous binary fluid, the fast growth
observed could thus be linked to the hydrodynamic modes expected in
any continuous description of Vicsek-like models~\cite{Toner1995,Toner_rev}.

\section{General discussion and outlook}
\label{discussion}

\subsection{Summary of main results}

We now summarize our main results before discussing 
them at a somewhat more general level.

We have provided ample evidence that the onset of collective motion in
Vicsek-style models is a discontinuous (first-order) phase transition,
with all expected hallmarks, in agreement with~\cite{Gregoire2004}. We
have made the (numerical) effort of showing this in the limits of
small and large velocity and/or density.

We have shown that the ordered phase is divided in two regions: near
the transition and down to rather low noise intensities, solitary
structures spanning the directions transverse to the global collective
velocity (the bands or sheets) appear, leading to an inhomogeneous
density field. For weaker noise, on the other hand, no such structures
appear, but strong, anomalous density fluctuations exist
and particles undergoes superdiffusive motion transverse to the mean velocity direction.

Finally, we have reported a linear growth (with time) of ordered domains 
when a disordered configuration is quenched in the ordered phase. This fast
growth can probably be linked to the expected emergence of 
long wavelength hydrodynamic modes
in the ordered phase of active polar particles models. 

\subsection{On the role of bands/sheets}

The high-density high-order traveling bands or sheets described here
appear central to our main findings.  They seem to be intimately
linked to the discontinuous character of the transition which can, to
some extent, be considered as the stability limit of these objects.
In the range of noise values where they are observed, the anomalous
density fluctuations present at lower noise intensities are
suppressed.

One may then wonder about the universality of these objects. Simple
variants of the Vicsek-style models studied here (e.g. restricting the
interactions to binary ones involving only the nearest neighbor) do
exhibit bands and sheets~\cite{Bertin}.  Moreover, the continuous
deterministic description derived by Bertin {\it et al.}~\cite{Bertin}
does possess localized, propagating solitary solutions rather similar
to bands~\cite{Bertin_prep}. Although the stability of these
solutions need to be further investigated, these results indicate that
these objects are robust and that their existence is guaranteed beyond
microscopic ``details''.  However, the emergence of regular, stable,
bands and sheets is obviously conditioned to the shape and the
boundary conditions of the domain in which the particles are allowed
to move. In rectangular domains with at least one periodic direction,
these objects can form, span across the whole domain, and move.  But
in, say, a circular domain with reflecting boundary conditions, they
cannot develop freely, being repeatedly ``frustrated''.  Nevertheless,
simulations performed in such a geometry indicate that the transition
remains discontinuous, with the ordered phase consisting of one or
several dense packets traveling along the circular boundary.  Note,
though, that these packets intermittently emit elongated structures
(bands) traveling towards the interior of the disk before colliding on
the boundary.  To sum up, bands appear as the ``natural'' objects in
the transition region, but they may be prevented by the boundaries to
develop into full-size straight objects.

At any rate, time-series of the order parameter such as the one
presented in Fig.~\ref{fig2} clearly show that the transition is
discontinuous irrespective of the geometry and boundaries of the
domain, and thus of whether bands/sheets can develop into stable
regular structures or not: the sudden, abrupt, jumps from the
disordered state to some ordered structure are tantamount to a
nucleation phenomenon characteristic of a discontinuous transition.

\subsection{A speculative picture}

We would now like to offer the following speculative general picture:
the key feature of the Vicsek-like models studied here ---as well as
of other models for active media made of self-propelled
particles~\cite{Csahok2002,BenJacob,Bertin,Topaz2004,Toner_rev}--- is
the coupling between density and order. Particles are forced to move,
and, since they carry informations about the order, advection, density
fluctuations and order are intimately linked. High-density means
strong local order (if noise is low enough) because the many particles
in a given neighborhood will adopt roughly the same orientation.  The
reverse is also true: in a highly-ordered region, particles will
remain together for a long time and thus will sweep many other
particles leading to a denser and denser group.

\begin{figure}
\includegraphics[width=8.6cm,clip]{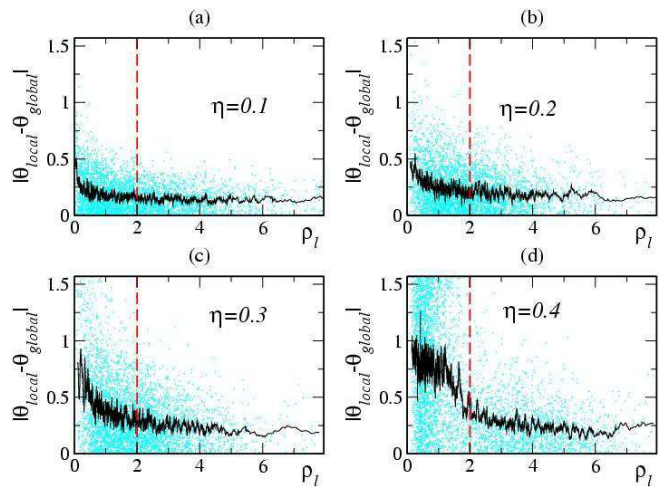}
\caption{(color online)
Scatter plots of local order parameter vs local density in the ordered phase
(angular noise, $\rho=2$, $v_{\rm 0}=0.5$ in a domain of size $1024 \times 256$, 
(same parameters as in Fig.~\ref{BandGNF}a). 
The local quantities were measured 
in boxes of linear size $\ell=8$. 
Here, the local order is represented by the
angle between the orientation $\Theta_{local}$ of the local order parameter 
and the global direction of motion $\Theta_{global}$. The black solid lines are running averages of the 
scatter plots. The red solid lines indicate the global density $\rho=2$ 
(and thus marks the percolation threshold in a two-dimensional square lattice).
(a,b): $\eta=0.1$ and 0.2: in the bandless regime, 
the ordered plateau starts below $\rho=2$, i.e. ordered regions percolate.
(c) approximately at the limit of existence of bands: the start of the plateau is near
$\rho=2$.
(d) at higher noise amplitude in the presence of bands.
}
\label{local-eq}
\end{figure}

At a given noise level, one can thus relate, 
in the spirit of some local equilibrium
hypothesis, local density to local order.
In practice, such an ``equation of state'' approach can be justified by 
looking, e.g., at a scatter plot of local order parameter vs local density.
Fig.~\ref{local-eq} reveals that, 
in the ordered bandless regime, such a scatter 
plot is characterized by a plateau over a large range of local density values
corresponding to order,
followed, below some crossover density, by more disordered local patches.
The regions in space where local density is below
this crossover level do not percolate in the bandless regime, 
and order can be maintained very steadily in the 
whole domain (this is corroborated by the fact that in spite of the large,
anomalous density fluctuations, the order parameter field is, 
on the other hand, rather constant, see Fig.~\ref{BandGNF}c).
The noise intensity at which bands emerge roughly corresponds to the value
where the low-density disordered regions percolate.
The reamining disconnected, dense patches then eventually self-organize 
into bands/sheets.
The emergence of these elongated structures is rather natural: moving packets
elongate spontaneously because they collect many particles;
superdiffusion in the directions perpendicular to the mean motion 
endow these nascent bands/sheets with some ``rigidity''. 
At still stronger noise, the bands/sheets are destroyed and global order
disappears. 

The above features are at the root of the approach by Toner and Tu 
\cite{Toner1995,Toner1998,Toner1998_2}. 
Their predictions of 
strong density fluctuations, transverse superdiffusion, 
and peculiar sound propagation
properties are correct as long as bands/sheets do not exist, i.e. for not too
strong noise intensities. 
This is indeed in agreement with their assumption that
the density field is statistically homogeneous in the frame moving
at the global velocity (albeit with strong fluctuations), which is only true
in the bandless regime.

\subsection{Outlook}

The results presented here are almost entirely numerical. Although
they were obtained with care, they need to be ultimately backed up by
more analytical results. A first step is the derivation of a
continuous description in terms of a density and a velocity field (or
some combination of the two), which would allow to go beyond
microscopic ``details''. In that respect the deterministic equation
derived by Bertin {\it et al.}  from a Boltzmann description in the
dilute limit~\cite{Bertin} is encouraging. However, one may suspect
that intrinsic fluctuations are crucial in the systems considered here
if only because some of the effective noise terms will be
multiplicative in the density. A mesoscopic, stochastic equation
description is thus a priori preferable.
This is especially true in view of the ``giant'' anomalous
fluctuations present in the bandless ordered phase.  These
fluctuations clearly deserve further investigation, all the more so as
they seem to be generic features of active particle
models~\cite{Mishra,Chate2006}.

Ongoing work is devoted to both these general issues.



\acknowledgements

Most of this research has been funded by the European Union 
via the FP6 StarFLAG project. Partial support from the French ANR
Morphoscale project is acknowledged.
We thank A. Vulpiani and M. Cencini for
introducing us to the method outlined in Ref. \cite{Vulpiani2000}.

\appendix
\section{Fluctuation-dissipation relation}

\begin{figure}
\includegraphics[width=8.6cm,clip]{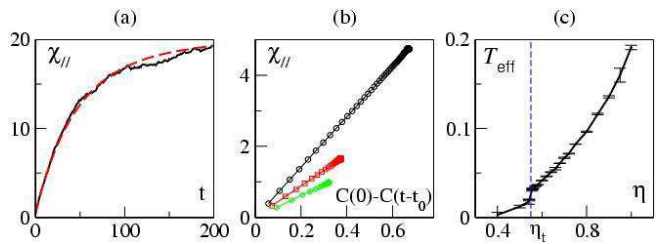}
\caption{(color online) Test of the fluctuation-dissipation relation
on the vectorial model with repulsive force ($\rho=2$, $v_{\rm 0}=0.3$,
$L=128$).  (a): susceptibility \em vs \em time at reduced noise
amplitude $\varepsilon=1 - \eta/\eta_{\rm t}=0.005$,
$|\vec{h}|=10^{-2}$(dashed red line) and $|\vec{h}|=10^{-3}$(plain black line). (b):
susceptibility \em vs. \em correlation at $|\vec{h}|=10^{-3}$ and
$\varepsilon=0.08$, $0.26$ and $0.44$ from top to bottom. (c):
effective temperature \em vs \em noise amplitude. The vertical dashed line marks the transition point.}
\label{FFDT}
\end{figure}

In~\cite{Czirok1997}, Vicsek {\it et al.} also studied the 
validity of the fluctuation-dissipation theorem and concluded,
from numerical analysis, that it is violated. Here we approach
this question again, using the ``dynamical'' approach put forward
in \cite{Cuku}, rather than the equilibrium used in \cite{Czirok1997}.
The fluctuation-dissipation relation is expressed as:

\begin{equation}
R(t-t_{\rm 0})=\frac{1}{T_{\rm eff}}\frac{\partial C(t-t_{\rm 0})}{\partial t}
\label{EFDT1}
\end{equation}
where $R$ is the response, $C$ the associated correlation function, and 
$T_{\rm eff}$ is some ``effective temperature''.

In both cases, at any rate, the key point is to investigate the
effect of an external field on the ordering process. Since here
one cannot rely on any Hamiltonian structure, the external
field remains somewhat arbitrary, as it cannot be unambiguously 
defined as the conjugate variable of the order parameter.

In~\cite{Czirok1997}, the field $\vec{h}$ was directly confronted to the
local average velocity, 
and the governing equation was replaced by:
\begin{equation}
\vec{v_i}(t+\Delta t) = v_{\rm 0} \,\, (\mathcal{R}_\eta\!\circ\vartheta)\left[
\sum_{j\in\mathcal{V}_i}\vec{v_j}(t) +\vec{h}\right] \;.
\label{EVfield}
\end{equation}
This manner of introducing $\vec{h}$ leads to an effective intensity
which depends on the local ordering: $\vec{h}$ is comparatively
stronger in disordered regions (an important effect at the early
stages of ordering) than in ordered regions. This could in fact
prevent the necessary linear regime to occur, even at very low field
values (see, for instance, Fig.~6 of \cite{Czirok1997}).  A mean-field
analysis has confirmed this view, showing a logarithmic variation of
the response with the field intensity~\footnote{Private communication
  from E.~Bertin.}.

To bypass this problem, we have preferred to use the following equation:
\begin{equation}
\vec{v_i}(t\!+\!\Delta t) = v_{\rm 0} \, \vartheta\!\left[
\!\sum_{j\in\mathcal{V}_i}\!\vec{v_j}(t)
\!+\!\left|\sum_{j\in\mathcal{V}_i}\!\vec{v_j}(t)\right| \vec{h}
+\eta \mathcal{N}_i \vec{\xi} \right] 
\label{EN2field}
\end{equation}
where the vectorial noise was chosen because, as shown above, it leads more 
easily to the asymptotic regime. The effective intensity of the field is now
proportional to the local order. 

Two (scalar) response functions can be defined in our problem: the longitudinal
response $R_\parallel$ along the field direction, and the transverse 
response $R_\perp$. We consider the former. In practice, we quenched, at time $t_{\rm 0}$, a strong-noise, 
highly-disordered system $\varphi(t_{\rm 0}) \approx 0$ to a smaller noise value and 
started applying the constant, homogeneous field $\vec{h}$ immediately. 
We then followed the
response of the system by monitoring the growth of the order parameter.
We measured the susceptibility $\chi_\parallel$
which is nothing but the integrated response function:
$\chi_{\parallel}(t,t_{\rm 0})=\int_{t_{\rm 0}}^{t}R_{\parallel}(t,t')dt'$.
In practice, we have:
\begin{equation}
\chi_{\parallel}(t-t_{\rm 0})=
\frac{1}{|\vec{h}|}\,\vec{\varphi}(t)\cdot \vartheta [\vec{h}] \;.
\end{equation}
In a well-behaved system, the susceptibility should be independent
of the amplitude of the field, at least at small enough values
(``linear'' regime). This is what we observed, as shown in 
Fig.~\ref{FFDT}a.

Correspondingly, the correlation function is defined as:
\begin{equation}
C(t-t_{\rm 0})=\frac{1}{v_{\rm 0}^2}\langle\vec{v}_i(t_{\rm 0})\cdot\vec{v}_i(t)\rangle_i \;.
\end{equation}
The fluctuation-dissipation relation (\ref{EFDT1}) can then be written in its 
integrated form:
\begin{equation}
\chi_{\parallel}(t-t_{\rm 0})=\frac{1}{T_{\rm eff}}
\left(C(0)-C(t-t_{\rm 0})\right)
\label{EFDT2}
\end{equation}
In Fig.~\ref{FFDT}b, we show that $\chi_{\parallel}$ and $C$ are related linearly
in time, confirming the validity of this relation and allowing an estimation
of $T_{\rm eff}$. 

This well-defined ---although not uniquely defined--- 
effective temperature varies as expected in parameter space. In particular,
it increases systematically with the noise strength $\eta$
(Fig.~\ref{FFDT}c), although this variation is not linear. Note also, that, 
intriguingly, there is a small jump of $T_{\rm eff}$ at the noise value 
corresponding to the transition in this case ($\eta_{\rm t}\simeq0.55$).


\begin{thebibliography}{10}

\bibitem{Pliny}
{Pliny the Elder},
\newblock Naturalis historia, 79.

\bibitem{Parrish1997}
J.~K. Parrish and W.~M. Hamner, editors,
\newblock {\em Animal Groups in Three Dimensions} (Cambridge University Press,
  Cambridge, 1997).

\bibitem{Albano}
E.~V. Albano,
\newblock Phys. Rev. Lett. {\bf 77}, 2129 (1996).

\bibitem{Couzin2003}
I.~D. Couzin and J.~Krause,
\newblock Advances in the Study of Behavior {\bf 32}, 1 (2003).

\bibitem{Czirok1997}
A.~Czir\'ok, H.~E. Stanley, and T.~Vicsek,
\newblock J. Phys. A {\bf 30}, 1375 (1997).

\bibitem{Gregoire2001}
G.~Gr\'egoire, H.~Chat\'e, and Y.~Tu,
\newblock Phys. Rev. E {\bf 64}, 011902 (2001).

\bibitem{Huepe}
C.~Huepe and M.~Aldana,
\newblock Phys. Rev. Lett. {\bf 92}, 168701 (2004).

\bibitem{Toner1995}
J.~Toner and Y.~Tu,
\newblock Phys. Rev. Lett. {\bf 75}, 4326 (1995).

\bibitem{Vicsek1995}
T.~Vicsek, A.~Czir\'ok, E.~Ben-Jacob, I.~Cohen, and O.~Shochet,
\newblock Phys. Rev. Lett. {\bf 75}, 1226 (1995).

\bibitem{Mermin}
N.~D. Mermin and H.~Wagner,
\newblock Phys. Rev. Lett. {\bf 17}, 1133 (1966).

\bibitem{Bertin}
E.~Bertin, M.~Droz, and G.~Gr\'egoire,
\newblock Phys. Rev. E {\bf 74}, 022101 (2006).

\bibitem{Birnir}
B.~Birnir,
\newblock J. Stat. Phys. {\bf 128}, 535 (2007).

\bibitem{Bussem}
H.~J. Bussemaker, A.~Deutsch, and E.~Geigant,
\newblock Phys. Rev. Lett. {\bf 78}, 5018 (1997).

\bibitem{Chate2006}
H.~Chat\'e, F.~Ginelli, and R.~Montagne,
\newblock Phys. Rev. Lett. {\bf 96}, 180602 (2006).

\bibitem{Couzin2002}
I.~D. Couzin,
\newblock J. Theor. Biol. {\bf 218}, 1 (2002).

\bibitem{Couzin2005}
I.~D. Couzin, J.~Krause, N.~Franks, and S.~Levin,
\newblock Nature {\bf 433}, 513 (2005).

\bibitem{Csahok1995}
Z.~Csah\'ok and T.~Vicsek,
\newblock Phys. Rev. E {\bf 52}, 5297 (1995).

\bibitem{Czirok1999}
A.~Czir\'ok, A.-L. Barab\'asi, and T.~Vicsek,
\newblock Phys. Rev. Lett. {\bf 82}, 209 (1999).

\bibitem{Czirok1999_2}
A.~Czir\'ok, M.~Vicsek, and T.~Vicsek,
\newblock Physica A {\bf 264}, 299 (1999).

\bibitem{Duparcmeur}
Y.~L. Duparcmeur, H.~Herrmann, and J.~P. Troadec,
\newblock J. Phys. I France {\bf 5}, 1119 (1995).

\bibitem{Comment_gg}
G.~Gr\'egoire, H.~Chat\'e, and Y.~Tu,
\newblock Phys. Rev. Lett. {\bf 86}, 556 (2001).

\bibitem{Gregoire2003}
G.~Gr\'egoire, H.~Chat\'e, and Y.~Tu,
\newblock Physica D {\bf 181}, 157 (2003).

\bibitem{Hemmingson}
J.~Hemmingsson,
\newblock J. Phys. A {\bf 28}, 4245 (1995).

\bibitem{Levine}
H.~Levine, W.-J. Rappel, and I.~Cohen,
\newblock Phys. Rev. E {\bf 63}, 017101 (2000).

\bibitem{Mikhailov}
A.~S. Mikhailov and D.~H. Zanette,
\newblock Phys. Rev. E {\bf 60}, 4571 (1999).

\bibitem{Mogilner}
A.~Mogilner and L.~Edelstein-Keshet,
\newblock J. Math. Biol. {\bf 38}, 534 (1999).

\bibitem{Oloan}
O.~J. O'Loan and M.~R. Evans,
\newblock J. Phys. A {\bf 32}, 99 (1999).

\bibitem{dOrsogna}
M.~R. d'Orsogna, Y.~Chuang, A.~Bertozzi, and L.~Chayes,
\newblock Phys. Rev. Lett. {\bf 96}, 104302 (2006).

\bibitem{Ken}
N.~Shimoyama, K.~Sugawara, T.~Mizuguchi, Y.~Hayakawa, and M.~Sano,
\newblock Phys. Rev. Lett. {\bf 76}, 3870 (1996).

\bibitem{Simha2002}
R.~A.~Simha and S.~Ramaswamy,
\newblock Physica A {\bf 306}, 262 (2002).

\bibitem{Simha2002_2}
R.~A.~Simha and S.~Ramaswamy,
\newblock Phys. Rev. Lett. {\bf 89}, 058101 (2002).

\bibitem{Szabo}
B.~Szab\'o {\em et~al.},
\newblock Phys. Rev. E {\bf 74}, 061908 (2006).

\bibitem{Toner1998}
J.~Toner and Y.~Tu,
\newblock Phys. Rev. E {\bf 58}, 4828 (1998).

\bibitem{Toner1998_2}
J.~Toner, Y.~Tu, and M.~Ulm,
\newblock Phys. Rev. Lett. {\bf 58}, 4828 (1998).

\bibitem{Topaz2004}
C.~M. Topaz and A.~L. Bertozzi,
\newblock SIAM J. Appl. Math. {\bf 65}, 152 (2004).

\bibitem{Topaz2006}
C.~Topaz, A.~Bertozzi, and L.~M.A.,
\newblock Bull. Math. Bio. {\bf 68}, 1601 (2006).

\bibitem{Vicsek1999}
T.~Vicsek, A.~Czir\'ok, I.~J. Farkas, and D.~Helbing,
\newblock Physica A {\bf 274}, 182 (1999).

\bibitem{Gregoire2004}
G.~Gr\'egoire and H.~Chat\'e,
\newblock Phys. Rev. Lett. {\bf 92}, 025702 (2004).

\bibitem{Nagy}
M.~Nagy, I.~Daruka, and T.~Vicsek,
\newblock Physica A {\bf 373}, 445 (2006).

\bibitem{Aldana2007}
M.~Aldana, {\it et al.},
\newblock Phys. Rev. Lett. {\bf 98}, 095702 (2007).

\bibitem{Lubeck}
S.~L\"ubeck,
\newblock Int. J. of Mod. Phys. B {\bf 18}, 3977 (2004).

\bibitem{Marcq}
P.~Marcq, H.~Chat\'e, and P.~Manneville,
\newblock Phys. Rev. Lett. {\bf 77}, 4003 (1996).

\bibitem{Marcq2}
P.~Marcq, H.~Chat\'e, and P.~Manneville,
\newblock Phys. Rev. E {\bf 55}, 2606 (1997).

\bibitem{Binder}
K.~Binder,
\newblock Monte carlo investigations of phase transitions and critical
  phenomena,
\newblock in {\em Phase transitions and critical phenomena}, edited by C.~Domb
  and M.~S. Green, Academic Press, 1976.

\bibitem{Privman}
V.~Privman, editor,
\newblock {\em Finite size scaling and numerical simulations of statistical
  systems} (ed. World scientific, Singapore, 1990).

\bibitem{Binder1997}
K.~Binder,
\newblock Rep. Prog. Phys. {\bf 60}, 487 (1997).

\bibitem{Efron}
B.~Efron,
\newblock {\em The Jackknife, The Bootstrap and other Resampling Plans} (SIAM,
  Philadelphia, 1982).

\bibitem{Zinn}
J.~C. Niel and J.~Zinn-Justin,
\newblock Nucl. Phys. B {\bf 280}, 355 (1987).

\bibitem{Borgs}
C.~Borgs and R.~Koteck\`y,
\newblock J Stat. Phys. {\bf 61}, 79 (1990).

\bibitem{Chate2007}
H.~Chat\'e, F.~Ginelli, and G.~Gr\'egoire,
\newblock \newblock Phys. Rev. Lett. {\bf 99},  229601 (2007).

\bibitem{Mishra}
S.~Mishra and S.~Ramaswamy,
\newblock Phys. Rev. Lett. {\bf 97}, 090602 (2006).

\bibitem{Narayan}
V.~Narayan, S.~Ramaswamy, and N.~Menon,
\newblock Science {\bf 317}, 105 (2007).

\bibitem{Ramaswamy}
S.~Ramaswamy, R.~A.~Simha, and J.~Toner,
\newblock Europhys. Lett. {\bf 62}, 196 (2002).

\bibitem{Vulpiani2000}
G.~Boffetta, A.~Celani, M.~Cencini, G.~Lacorata, and A.~Vulpiani,
\newblock Chaos {\bf 10}, 50 (2000).

\bibitem{Hohenberg}
P.~C. Hohenberg and B.~I. Halperin,
\newblock Rev. Mod. Phys. {\bf 49}, 435 (1977).

\bibitem{Toner_rev}
J.~Toner, Y.~Tu, and S.~Ramaswamy,
\newblock Annals Of Physics {\bf 318}, 170 (2005).

\bibitem{Bertin_prep}
E.~Bertin, M.~Droz, and G.~Gr\'egoire,
\newblock Hydrodynamic equations for self-propelled particles: a derivation
  from the microscopic dynamics,
\newblock in preparation, 2007.

\bibitem{Csahok2002}
Z.~Csah\`ok and A.~Czir\`ok,
\newblock Physica A {\bf 243}, 304 (2002).

\bibitem{BenJacob}
E.~Ben-Jacob, I.~Cohen, and H.~Levine,
\newblock Adv. in Phys. {\bf 49}, 395 (2000).

\bibitem{Cuku}
L.~F. Cugliandolo, J.~Kurchan, and L.~Peliti,
\newblock Phys. Rev. E {\bf 55}, 3898 (1997).

\end{thebibliography}

\end{document}